\documentclass[prd,aps,showpacs,twocolumn,nofootinbib]{revtex4-1}

\usepackage{amsmath,amssymb,amsthm,wasysym}
\usepackage{graphicx}
\usepackage{psfrag}
\usepackage{epstopdf}
\usepackage{color}
\usepackage{mathrsfs}
\usepackage{fixmath}

\newcommand{\bse}{\begin{subequations}}

\DeclareSymbolFontAlphabet{\mathrsfs}{rsfs}
\DeclareMathAlphabet\mathbfcal{OMS}{cmsy}{b}{n}

\def\lm{{\ell m}}

\def\de{\partial}
\def\lm{{\ell m}}

\def\ii{{\rm i}}

\def\r{{\hat{r}}}

\def\F{\hat{F}}

\def\O{{\cal O}}

\def\x{\hat{\hat{x}}}

\newcommand{\be}{\begin{equation}}
\newcommand{\ee}{\end{equation}}

\usepackage{float}
\definecolor{cyan}{rgb}{0,0.9,0.9}
\definecolor{orange}{rgb}{0.9,0.5,0}
\definecolor{magenta}{rgb}{1,0,1}
\definecolor{purple}{rgb}{0.8,0.4,0.8}

\begin{document}
\title{Horizon-absorbed energy flux in circularized, nonspinning  black-hole binaries and its effective-one-body representation}
\author{Alessandro \surname{Nagar}}
\affiliation{Institut des Hautes Etudes Scientifiques, 91440 Bures-sur-Yvette, France}

\author{Sarp \surname{Akcay}}
\affiliation{School of Mathematics, University of Southampton, Southampton SO17 1BJ, UK}

\begin{abstract}
We propose, within the effective one body (EOB) approach, a new, resummed, analytical representation 
of the gravitational wave energy flux absorbed by a system of two circularized (nonspinning) black holes.
This expression is such to be well-behaved in the strong-field, fast motion regime, notably up to
the EOB-defined last unstable orbit.
Building conceptually upon the procedure adopted to resum the multipolar asymptotic energy flux, 
we introduce a {\it multiplicative} decomposition of the multipolar absorbed flux made by three factors: 
(i) the leading-order contribution, (ii) an ``effective source'' and (iii) a new residual amplitude 
correction $(\tilde{\rho}_\lm^H)^{2\ell}$. In the test-mass limit, we use a frequency-domain perturbative 
approach to accurately compute numerically  the horizon-absorbed fluxes along a sequence of 
stable and unstable circular orbits and we extract from them the functions $\tilde{\rho}_\lm^H$. These 
quantities are then fitted via rational functions. The resulting analytically represented test-mass
knowledge is then suitably {\it hybridized} with lower-order analytical information that is valid for 
any mass ratio. This yields a resummed representation of the absorbed flux for a generic, circularized, 
nonspinning black-hole binary. Our result adds new information to the state-of-the-art calculation of 
the absorbed flux at fractional 5~post-Newtonian order [S.~Taylor and E.~Poisson, Phys. Rev. D {\bf 78} 084016 (2008)], 
that is recovered in the weak-field limit approximation by  construction.
\end{abstract}
\date{\today}


\maketitle

\section{Introduction}
A ground-based network of interferometric gravitational wave (GW) detectors is currently being
upgraded to an improved sensitivity and is expected to detect, within a few years, the GW signal
emitted during the inspiral and merger of compact binaries.
The effective one body (EOB) approach to the general relativistic two-body
dynamics~\cite{Buonanno:1998gg,Buonanno:2000ef,Damour:2000we,Damour:2001tu,Damour:2008gu,
Damour:2008qf,Damour:2009wj,Damour:2009ic,Barausse:2009xi,Pan:2010hz,Barausse:2011ys,Barausse:2011dq}
is the most promising analytical tool currently available to model the dynamics
and radiation emitted by coalescing compact binaries, notably binary black holes (BBH).
The EOB analytical description of the binary dynamics mainly relies on two building blocks: the resummed
EOB Hamiltonian $H_{\rm real}$, that takes into account conservative effects; and the mechanical
angular momentum loss ${\cal F}_\varphi$, also given in resummed form, that takes into account
nonconservative effects due to GW emission (radiation reaction).
The current analytical construction of both $H_{\rm real}$  and ${\cal F}_{\varphi}$ has reached a high
degree of sophistication for both spinning and nonspinning binaries. Several recent comparisons with
numerical relativity simulations~\cite{Pan:2009wj,Baiotti:2010xh,Baiotti:2011am,Pan:2011gk,Damour:2011fu}
of inspiralling and coalescing relativistic binaries, and with gravitational self force
calculations~\cite{Barack:2010ny} have vindicated the EOB approach both as a mean to deeply
understand the physical outcome of such simulations and as a practical tool to build accurate
analytical template waveforms for data analysis purposes~\cite{Damour:2010zb}.

In the typical EOB implementation for circularized black hole binaries with comparable masses, the
radiation reaction force ${\cal F}_\varphi$ is   computed only from the {\it asymptotic} GW energy flux.
This choice is typically rather accurate during the early inspiral phase. On the other hand, the cumulative
effect during the late inspiral and the plunge --- typically after the crossing of the last stable orbit
(LSO) --- of horizon absorption can have a nonnegligible, though small, influence on the phasing. This might
have some relevance (especially in the spinning case) in the ongoing effort in interfacing numerical
simulations with analytical, EOB-based, predictions.

The analytical results about horizon absorption of GWs are collected
in a few papers: in a pioneering study in the test-mass limit, Poisson and Sasaki~\cite{Poisson:1994yf}
computed analytically the leading-order contribution to black-hole absorption for a particle on
circular orbits around a Schwarzschild black hole, providing closed analytical formulas valid 
for all multipoles. This work was then generalized by Tagoshi, Mano and Takasugi~\cite{Tagoshi:1997jy},
who considered black-hole absorption for a particle in circular equatorial orbits around a Kerr black hole.
Alvi~\cite{Alvi:2001mx} provided the leading order calculation of the growth of mass
and angular momentum due to horizon absorption for a general, spinning binary.

More recently, Poisson and coworkers have been very active in investigating the dynamics of
tidally deformed black holes~\cite{Poisson:2005pi,Taylor:2008xy,Comeau:2009bz,Poisson:2009qj},
with several computations (in the small-hole/slow-motion approximation) of the tidal heating
and torquing of a black hole induced by an external tidal environment. In their language, the
expression``tidal heating'' of a black hole refers to the change of its mass that occurs as a
result of the tidal interaction; ``tidal torquing'' refers to the change of its angular momentum.
 The state-of-the-art analytical information that we shall exploit here (in the nonspinning case only)
is the tidal heating of a black hole of mass $m_A$ created by an external body of mass
$m_B$ as computed by Taylor and Poisson~\cite{Taylor:2008xy} for a circularized, spinning,
binary in the weak-field approximation. The result of Ref.~\cite{Taylor:2008xy}
is obtained at a higher order of accuracy than what is achieved in Ref.~\cite{Alvi:2001mx}.
The crucial physical information that can be extracted from these analytical studies
can be summarized in two points: absorption effects are fractionally more important
(i) when the black holes are spinning and (ii) when the mass-ratio is large.

In the comparable mass case, and building on a suggestion of Price and Whelan~\cite{Price:2001un},
the influence of absorption effects on the phasing was  also investigated by
Alvi~\cite{Alvi:2001mx} for the first time, using the weak-field approximation
and at leading order. The conclusion of this study is that horizon absorption
would change the phasing by no more than a tenth of a cycle by the time the binary merge.
Notably, this leading-order estimate is much larger than the uncertainties of state-of-the-art
numerical simulations of coalescing black hole binaries~\cite{Scheel:2008rj,Reisswig:2009us},
suggesting that it can be an effect extractable from the numerical data.
Consistently with this understanding, a recent NR simulation
of a nearly-extremal spinning black hole binary~\cite{Lovelace:2011nu} has shown a remarkable
agreement, during the early inspiral, between the rate of growth of the irreducible mass and
angular momentum of the black hole horizons computed numerically and Alvi's analytical prediction.
This numerical result suggests that it is recommendable to include horizon absorption effects in
the EOB-based analytical modeling of coalescing black hole binaries.
To our knowledge, absorption terms have been included in the EOB model (though in nonresummed
form) within the context of an extreme-mass-ratio (spinning) inspiralling binary,
a study that aimed at determining the suitability of EOB-based template waveforms for
LISA data analysis purposes~\cite{Yunes:2010zj,Yunes:2009ef}.

For noncircularized binaries the effect of black-hole absorption is more
relevant~\cite{Hughes:2001jr,Martel:2003jj,Barack:2010tm}.
For example, in the test-mass limit (on Schwarzschild)
absorption effects were thoroughly investigated by Martel~\cite{Martel:2003jj} (and more
recently in Ref.~\cite{Vega:2011ue}): he considered both eccentric and parabolic orbits and found
that the absorbed flux is nonneglible (larger than $1\%$ of the asymptotic flux) 
for orbits whose periastron is smaller than $5M$.
On the basis of these test-mass results, one expects that horizon fluxes should have some relevance
(even in the nonspinning case)  also for comparable mass binaries with an analogous dynamical
setup, for example when a zoom-whirl behavior is present~\cite{Pretorius:2007jn,Healy:2009zm,Sperhake:2009jz,Sperhake:2010uv}.

The main purpose of this paper is to provide an analytical expression of the (multipolar) horizon-absorbed
flux for circularized, nonspinning, binaries within the EOB framework that is valid also in the strong-field,
high-velocity regime, i.e. possibly up to the EOB-defined last unstable orbit.
This result builds upon three different pieces of knowledge: (i) the leading-order and next-to-leading
order analytical results for the multipolar horizon flux of Refs.~\cite{Poisson:1994yf,Taylor:2008xy};
(ii) the accurate numerical computation of the horizon fluxes for a test-particle orbiting along 
stable and unstable circular orbits of a Schwarzschild black-hole; (iii) the idea of suitably factorizing
the analytical representation of the horizon flux, analogously to what was proposed
(actually, at the waveform level) in Refs.~\cite{Damour:2008gu,Damour:2009kr}. In addition, the numerical
computation of the horizon-absorbed flux for unstable circular orbits of Schwarzschild is presented
here for the first time.

The paper is organized as follows: Sec.~\ref{sec:PN} reviews the next-to-leading order
analytical results of Refs.~\cite{Poisson:1994yf,Taylor:2008xy}. In Sec.~\ref{sec:abs_flux} we
first present the structure of the factorized resummation of the asymptotic multipolar energy flux for
circularized binaries within the EOB approach and then we introduce our, conceptually analogous,
new, resummed representation of the multipolar horizon flux. In Sec.~\ref{sec:test_mass} we
calculate the residual amplitude corrections $\rho_\lm^H$ from the numerically computed horizon
flux in the test-mass limit and in Sec.~\ref{sec:horizon_nu} we show how to use them together
with the analytical results of Sec.~\ref{sec:abs_flux} so as to obtain an explicit analytical
expression of the horizon flux that holds for any mass ratio. We collect our
conclusions in Sec.~\ref{sec:end}.
We use geometrized units with $G=c=1$.

\section{Absorbed energy flux: PN-expanded analytical results}
\label{sec:PN}
We start by reviewing the analytical results of Refs.~\cite{Poisson:1994yf,Taylor:2008xy}.
We consider a system made of two nonspinning black holes of masses $m_A$ and $m_B$
(with total mass $M=m_A+m_B$) in circular orbit.  In the following we
shall use  the symmetric mass ratio parameter $\nu=\mu/M$, where $\mu=m_A m_B/M$.
This parameter varies between $\nu=0$ (test-mass) and $\nu=1/4$ (equal-mass).
Here, the ``test-mass'' limit is defined by the condition
$m_B\ll m_A$, i.e. $M\simeq m_A$ becomes the mass of the larger black-hole,
$\mu\simeq m_B$ that of the smaller one and $\nu\simeq m_B/m_A$.
The total energy flux emitted by the system can be written as a multipolar
expansion of the form
\be
F^{(\ell_{\rm max})} = \sum_{\ell=2}^{\ell_{\rm max}}\sum_{m=1}^{\ell} F_\lm ,
\ee
where each multipolar contribution\footnote{Our definition assumes that the $+m$ and $-m$ modes are summed together.} 
$F_\lm$ is given by the sum of the flux emitted at infinity, $F_{\ell m}^{\infty}$ and the ``horizon'' absorbed
flux, $F_{\ell m}^{H}$,
\begin{equation}
\label{eq:sum_fluxes}
F_{\ell m}=F_{\ell m}^{\infty}+F_{\ell m}^{H}.
\end{equation}
In the test-particle limit the absorbed flux is the energy flux through the event horizon of
the large mass. In the case of two Schwarzschild black holes,
it is the sum of the fluxes through the two horizons\footnote{Note
that our definition of $F^H_\lm$ in the test-mass limit is neglecting
the flux absorbed by the black hole of smaller mass that
is in any case smaller by the fourth-power of the mass ratio.
It is however remarkable that the quadrupolar contribution to
this quantity has been explicitly computed by Poisson in closed
form, see Eq.~(8.52) of Ref.~\cite{Poisson:2004cw}.
If the need arises, it can just be added to Eq.~\eqref{eq:sum_fluxes}.}.

For a test-mass on Schwarzschild circular orbits the functions $F_{\lm}^{\infty}$
have been recently computed by Fujita~\cite{Fujita:2011zk} up to the 14
post-Newtonian (PN) order. This achievement builds upon a
systematic method that treats black-hole perturbations~\cite{Mano:1996vt}
and it improves on previous state-of-the-art 5.5PN-accurate
calculation~\cite{Tanaka:1997dj}.
For the general $\nu$-dependent case, the PN knowledge of $F_\lm^\infty$
has been pushed up to 3.5PN order~\cite{Blanchet:2001aw,Blanchet:2001ax,Blanchet:2004ek,Blanchet:2005tk}.

As mentioned in the introduction, much less analytical information
is available for $F_\lm^H$. The leading order (LO) contributions (for each
multipole) in the test-particle limit (with the {\it mass-ratio} $\mu/M=\nu\ll 1$)
were computed long ago in closed form by Poisson and Sasaki~\cite{Poisson:1994yf}.
Combining their Eqs.~(5.7) and (5.13)-(5.18), one can rewrite each
multipolar contribution in compact form as
\begin{align}
\label{eq:PS94}
F_\lm^{(H_{\rm LO},\epsilon)}(x) = \nu^2 {\cal N}_{\lm}^{{H},(\epsilon)}x^{2\ell+5+\epsilon}\left[{}_{-\epsilon}Y_\lm\left(\dfrac{\pi}{2},0\right)\right]^2
\end{align}
where $x=(M\Omega)^{2/3}$ is the PN frequency parameter ($\Omega$ is the system orbital frequency)
and $\epsilon$ denotes the parity of the flux, i.e., even ($\epsilon=0$) for
mass-generated multipoles and odd ($\epsilon=1$) for current-generated ones.
In the circular case, $\epsilon$ is equal to the parity of the sum
$\ell+m:\,\epsilon=\pi(\ell+m)$, that is $\epsilon=0$ when $\ell+m$ is even
and $\epsilon=1$ when $\ell+m$ is odd.
The quantities ${}_{0}Y_\lm$ and ${}_{-1}Y_\lm$ are the spherical harmonics of spin-weight $s=0$
and $s=-1$, that are given explicitly in Eqs.~(5.7) and (5.13) of Ref.~\cite{Poisson:1994yf},
while
\begin{align}
{\cal N}_\lm^{{H},(0)} & = 32\pi \left[\dfrac{m\:\ell!}{(2\ell+1)!!}\right]^2\dfrac{(\ell+1)(\ell+2)}{\ell(\ell-1)},\\
{\cal N}_\lm^{{H},(1)} & = 128\pi\left[\dfrac{m(\ell+1)!}{\ell(2\ell+1)!!}\right]^2\dfrac{\ell+2}{\ell-1}.
\end{align}

When the masses are comparable (and possibly equal), the state-of-the-art
analytical knowledge is due to Taylor and Poisson~\cite{Taylor:2008xy}.
As in the test-mass limit, the calculation is based on the weak-field
approximation but, contrary to the result of Ref.~\cite{Poisson:1994yf},
it is complete at next-to-leading order (NLO), i.e. 1PN fractional accuracy.

Let us now review the results Ref.~\cite{Taylor:2008xy} and recast
them in a form that is useful for our purpose.
Focusing on the object labeled by $A$, the NLO (i.e., 10PN accurate)
``rate at which the black hole acquires mass by tidal heating''
is expressed as the sum of a $\ell=m=2$ and a $\ell=2$, $m=1$
contributions, which explicitly read (see Eq.~(9.1)
of~\cite{Taylor:2008xy})
\begin{align}
&\dot{m}^{22}_A = \dfrac{16}{45}m_A^6 {\dot{\bar{\cal E}}}_{ab}{\dot{\bar{\cal E}}}^{ab}\nonumber\\
               &=\dfrac{32}{5}\dfrac{m_A^6 m_B^2}{M^8}\left(\dfrac{M}{\r}\right)^9\left(1-\dfrac{6 m_A^2 + 14 m_A m_B + 7 m_B^2}{M^2}\dfrac{M}{\r}\right),\\
&\dot{m}^{21}_A = \dfrac{16}{45}m_A^6 {\dot{\bar{\cal B}}}_{ab}{\dot{\bar{\cal B}}}^{ab}=\dfrac{32}{5}\dfrac{m_A^6 m_B^2}{M^8}\left(\dfrac{M}{\r}\right)^{10} ,
\end{align}
where $\r$ is the orbital separation in harmonic coordinates.
The quantities  $\bar{\cal E}_{ab}$ and  $\bar{\cal B}_{ab}$ are the electric and magnetic
tidal  moments in the black hole frame and the overdot refers to
time derivative in the same frame; we used the relations mentioned
after Eqs.~(9.2)-(9.3) of~\cite{Taylor:2008xy}) to write explicitly the right-hand sides.
Summing together these two pieces gives the final, fractionally 1PN-accurate,
contribution to the expression for tidal heating given by Eq.~(9.4)
of Ref.~\cite{Taylor:2008xy}.
The total multipolar horizon fluxes of the binary system are then given by
\begin{align}
F^H_{22}(\r;\,\nu)&=\dot{m}_A^{22} + \dot{m}_B^{22},\\
F^H_{21}(\r;\,\nu)&=\dot{m}_A^{21} + \dot{m}_B^{21},
\end{align}
which become
\begin{align}
F^{H}_{22}(\r;\,\nu) &= \dfrac{32}{5}\left(\dfrac{M}{\r}\right)^9\nu^2\left(1-4\nu + 2\nu^2\right)\nonumber\\
                 &\times \left(1-\dfrac{6-22\nu + 5\nu^2 + 2\nu^3}{1-4\nu+2\nu^2}\dfrac{M}{\r}\right),\\
F^{H}_{21}(\r;\,\nu) &= \dfrac{32}{5}\left(\dfrac{M}{\r}\right)^{10}\nu^2\left(1-4\nu+2\nu^2\right),
\end{align}
where we made now explicit the dependence on the symmetric mass ratio parameter $\nu$. One then
wants to express the harmonic radius $\r$ in terms of the gauge-invariant frequency parameter
$x= (M\Omega)^{2/3}$. At the needed 1PN accuracy, this relation reads~\cite{Blanchet:2006zz}
\be
\dfrac{M}{\r}=x\left[1+\left(1-\dfrac{\nu}{3}\right)x+\O(x^2)\right],
\ee
so to get the gauge-invariant results
\begin{align}
\label{eq:F22Hpn}
F_{22}^{H}(x;\,\nu)&=\dfrac{32}{5}x^9 \nu^2(1-4\nu + 2\nu^2)\nonumber\\
 &\times \left(1+\dfrac{3-17\nu+25\nu^2-8\nu^3}{1-4\nu+2\nu^2}x+\O(x^2)\right),\\
\label{eq:F21Hpn}
F_{21}^H(x;\,\nu)&=\dfrac{32}{5}x^{10}\nu^2\left(1-4\nu+2\nu^2\right)\left(1+\O(x)\right).
\end{align}
Note that the leading-order contributions
\begin{align}
\label{eq:F22LO}
F_{22}^{(H_{\rm LO},0)}(x;\,\nu)&=\dfrac{32}{5}x^9 \nu^2(1-4\nu + 2\nu^2),\\
\label{eq:F21LO}
F_{21}^{(H_{\rm LO},1)}(x;\,\nu)&=\dfrac{32}{5}x^{10}\nu^2\left(1-4\nu+2\nu^2\right),
\end{align}
reduce to Eq.~\eqref{eq:PS94} (for $\ell=2$, $m=1,2$) in the $\nu\to 0$ limit.
Equations~\eqref{eq:F22Hpn}-\eqref{eq:F21Hpn} are basically the only analytical
information present in the literature about horizon absorption for nonspinning
black-hole binaries in the comparable mass case\footnote{Notably, in Ref.~\cite{Poisson:2005pi}
Poisson was able to compute also the octupole contributions to the metric of a
distorted black hole. This means that in the literature there is already
enough information to compute at leading order the $\ell=3$ black-hole absorption
part by following the same procedure of Ref.~\cite{Taylor:2008xy}.}.

\section{Resummed representation of the absorbed energy flux}
\label{sec:abs_flux}

The purpose of this section is to obtain a physically motivated analytical
expression of the absorbed energy flux for circularized, black-hole binaries
(for any mass ratio $\nu$) that has the property to remain valid also away
from the weak-field, slow-motion assumption that is at the basis of the
derivation of Eqs.~\eqref{eq:F22Hpn}-\eqref{eq:F21Hpn}.
More precisely, we would like to obtain, by a physically motivated \emph{guess}
within the EOB framework, a {\it resummed} analytical expression that is
well behaved in the strong-field, fast-motion regime, possibly up to the
last {\it unstable} circular orbit.
To to so, we will introduce a factorized, analytical model
of the absorbed multipolar flux that is inspired by the procedure followed
in Ref.~\cite{Damour:2008gu} to treat the analogous problem in the case of
asymptotic fluxes.

\subsection{Factorization of the multipolar asymptotic flux}
Let us first recall the prescription that is used to compute the
resummed asymptotic energy flux (in a factorized form) for circularized
binaries in the EOB approach.
For circular orbits, the dynamics is typically parameterized in terms of
the gauge-invariant frequency parameter $x=(M\Omega)^{2/3}$ introduced above.
The conservative dynamics is described by the real EOB Hamiltonian
\be
H_{\rm real}=M\sqrt{1+2\nu\left(\hat{H}_{\rm eff}-1\right)},
\ee
where $\hat{H}_{\rm eff}=H_{\rm eff}/\mu$ is the (reduced) effective
Hamiltonian. When $\nu\to 0$, $\hat{H}_{\rm eff}$
reduces to the usual conserved energy of a test-mass $\mu$
in a Schwarzschild background of mass $M$. The explicit expression
of $\hat{H}_{\rm eff}$ as a function of the frequency parameter
$x$ cannot be written in closed form, but rather in parameterized
form in terms of the EOB inverse radius parameter $u\equiv M/r$,
that is as
\be
\label{eq:Heff}
\hat{H}_{\rm eff}=\sqrt{A(u)(1+j^2 u^2)},
\ee
where $A(u)$ is the EOB radial potential and $j$ is the dimensionless
angular momentum along circular orbits.
The PN-expansion of $A(u)$ at 3PN accuracy is
\be
A^{\rm Taylor}=1 - 2u + 2\nu u^3 + \left(\dfrac{94}{3}-\dfrac{41}{32}\pi^2\right)\nu u^4 + \O(u^5),
\ee
but then we use it in a Pad\'e resummed form, that allows for the presence
of an ``effective'' horizon, an effective ``light-ring'' (i.e., corresponding to
the last {\it unstable} orbit) and of the last stable orbit (LSO).
Since we work at 3PN accuracy, we use a (1,3) Pad\'e representation
\be
A(u) = P^1_3\left[A^{\rm Taylor}\right]=\dfrac{n_0+n_1 u}{d_0 + d_1 u + d_2 u^2 +d_3 u^3}
\ee
The circular orbits in the EOB formalism are determined by the condition
$\de_u\{A(u)[1+j^2 u^2]\}=0$ which leads to the following parametric representation
of the squared angular momentum
\be
j^2(u)=-\dfrac{A'(u)}{(u^2 A(u))'},
\ee
where the prime stands for $d/du$. Inserting this $u$-parametric representation
of $j^2$ in Eq.~\eqref{eq:Heff} defines the $u$-parametric representation of the
effective Hamiltonian $\hat{H}_{\rm eff}(u)$. One can then obtain $\hat{H}_{\rm eff}$
as a function of $x$ by eliminating $u$ between $\hat{H}_{\rm eff}(u)$ and the corresponding
$u$-parametric representation of the frequency parameter $x=(M\Omega)^{2/3}$ obtained
by the angular Hamilton equation of motion in the circular case
\be
M\Omega(u) = \dfrac{1}{\mu}\dfrac{\de H_{\rm real}}{\de j}=\dfrac{MA(u)j(u)u^2}{H_{\rm real}\hat{H}_{\rm eff}}.
\ee
The use of the variable $x$ to parameterize the adiabatic circular dynamics is the standard choice
when one focuses on {\it stable} circular orbits~\cite{Damour:2000we,Damour:2008gu} (i.e., orbits corresponding
to the local minimum of the effective potential $A(u)(1+j^2u^2)$~\cite{Buonanno:1998gg,Damour:2004bz}
for a given value of $j$). In this case, the factorization of the waveform introduced in~\cite{Damour:2008gu}
allows us to factorize each multipole of the asymptotic energy flux as
\be
\label{eq:flux_std}
F_{\ell m}^{(\infty,\epsilon)}(x;\,\nu) = F_\lm^{(N,\epsilon)}(x;\,\nu) \hat{F}^{(\infty,\epsilon)}_\lm(x;\,\nu),
\ee
where $F_\lm^{(N,\epsilon)}(x;\,\nu)$ is the Newtonian contribution for circular orbits,
$\hat{F}^{(\infty,\epsilon)}_\lm(x;\,\nu)$ symbolically represents all higher-order
relativistic corrections and $\epsilon$ is the parity of the sum $\ell+m$.
To be precise, the quantities $F_\lm^{(N,\epsilon)}(x;\,\nu)\propto x^{3+\ell+\epsilon}$
are the fluxes computed from the Newtonian contribution to the circular
gravitational waveform, $h_\lm^{(N,\epsilon)}$ as given by Eq.~(4) of Ref.~\cite{Damour:2008gu}.
The relativistic correction factor is written in resummed form as
\be
\label{eq:hatF}
\hat{F}^{(\infty,\epsilon)}_\lm(x;\,\nu) = \left(\hat{S}_{\rm eff}^{(\epsilon)}(x)\right)^2 |T_\lm(x)|^2 \left(\rho^\infty_\lm(x;\,\nu)\right)^{2\ell},
\ee
where $S_{\rm eff}^{(\epsilon)}$ is the effective
source of the field, 
\begin{align}
\hat{S}^{(0)}_{\rm eff}(x)&\equiv \hat{H}_{\rm eff}(x) \qquad\qquad\,\;\; \ell+m\quad{\rm even}, \\
\hat{S}^{(1)}_{\rm eff}(x)&\equiv \hat{j}(x)=x^{1/2}j(x) \quad \ell+m\quad{\rm odd} .
\end{align}
The quantity $T_\lm$ is the ``tail factor'' that takes into account the infinite number of
``leading logarithms'' entering the transfer function between the near zone multipolar wave
and the far-zone one due to {\it tail} effects. This factor is written as~\cite{Damour:2008gu}
\begin{align}
\label{eq:tail}
  T_\lm(x) = \dfrac{{\rm \Gamma}(\ell+1-2\ii\hat{\hat{k}})}{{\rm \Gamma}(\ell+1)}e^{\pi\hat{\hat{k}}}e^{2\ii \hat{\hat{k}}\log(2 k r_0)},
\end{align}
where $\hat{\hat{k}}=m(H_{\rm real}/M )x^{3/2}$, $k=(m/M) x^{3/2}$ and $r_0=2/\sqrt{e}$~\cite{Fujita:2010xj}.
Finally, $\rho_\lm^\infty$ are the residual amplitude corrections that are essentially
given as Taylor series (modulo $log(x)^n$ terms) of the form 
$\rho^\infty_\lm(x;\,\nu)=1 + x + x^2 + x^3 +\dots$ with
$\nu$-dependent coefficients. In the test-mass limit, the $\rho_\lm$'s are currently
analytically known  at 14PN order~\cite{Fujita:2011zk}. In the general, $\nu$-dependent
case  the available analytical information is more limited, nonetheless  it gets up
to 3PN order~\cite{Damour:2008gu}. This $\nu$-dependent information is then ``hybridized''
with the test-mass limit results (at 5PN-accuracy only) to compute the so-called
$3^{+2}$PN approximation that has been used recently in several comparisons
between EOB predictions and numerical simulations of nonspinning
inspiralling and coalescing compact binaries~\cite{Damour:2009kr,Lackey:2011vz,Baiotti:2011am,Pan:2011gk,Damour:2011fu}.
Moreover, in the test-mass limit the $\rho_\lm^\infty$'s have also been computed numerically,
using black-hole perturbation theory, for stable~\cite{Damour:2008gu} and
unstable~\cite{Bernuzzi:2011aj} circular orbits.

When one wants to consider also the {\it unstable} circular orbits predicted by the EOB
conservative dynamics~\cite{Buonanno:1998gg,Damour:2004bz} (i.e., orbits corresponding
to the local maximum of the effective potential $A(u)(1+j^2u^2)$ for a given value of $j$),
as we are going to do here, a complication occurs, since the variable $x$
is found to be ill-behaved for these orbits. When expressed as a function of $u$, $x$ does not
grow monotonically (as in the test-mass limit, where it is simply $x=u$), but it has a maximum
and then decreases, to go to zero at the EOB last {\it unstable} orbit (or EOB-defined ``light-ring'',
that in the adiabatic approximation is defined as the maximum of the function $u^2 A(u)$).
The effect of this on the fluxes $\hat{F}^{(\infty,\epsilon)}_\lm(x;\,\nu)$ is that they grow monotonically 
until the maximum of $x$ and then turn back (i.e., they are not functions anymore). This  behavior is
obviously qualitatively incompatible with the test-mass result, where the asymptotic
fluxes are always growing monotonically until they diverge at the light-ring
position, $x=1/3$.

To meaningfully compute a resummed representation of the energy flux along unstable
orbits we need first to use a different, well-defined, frequency parameter.
A simple possibility is given by the following variable
\be
\label{eq:xbar_of_x}
\x = (H_{\rm real}\Omega)^{2/3}=x \left(\dfrac{H_{\rm real}}{M}\right)^{2/3},
\ee
which does not vanish at the EOB light-ring as a function of $u$ and it is continuously
connected to the standard frequency parameter $x$ both in the test-mass and Newtonian
limits. The notation $\x$ that we use here reminds us of the variable $\hat{\hat{k}}$
in Eq.~\eqref{eq:tail}, that can be written as $\hat{\hat k} = m\x^{3/2}$.
More importantly, the use of $\x$ suggests a slightly different factorization of the resummed
energy flux. To derive it, we take Eq.~\eqref{eq:flux_std} and we first make it explicit the
dependence of $x$ on $\x$, i.e.
\begin{align}
\label{eq:Finfty}
F^{(\infty,\epsilon)}_\lm\left(x(\x);\,\nu\right) &= F_\lm^{(N,\epsilon)}\left(x(\x);\,\nu\right)\nonumber\\
                     & \times\left(\hat{S}_{\rm eff}^{(\epsilon)}(x(\x))\right)^2 |T_\lm(\x)|^2 \left(\rho_\lm^\infty(x(\x);\,\nu)\right)^{2\ell},
\end{align}
where $x(\x)$ symbolically represents the inverse of Eq.~\eqref{eq:xbar_of_x}.
Note that in the tail factor the dependence on $\x$ is left explicit
because of Eq.~\eqref{eq:tail}. We then replace the function $x(\x)$ in
$F_\lm^{(N,\epsilon)}\left(x(\x);\,\nu\right)$ and $\rho_\lm^\infty(x(\x);\,\nu)$ by
\begin{align}
\label{eq:xofxhat}
&\bar{x}_{\rm 3PN}(\x) \equiv \x f(\x)\nonumber\\
   &=\x\bigg[1 + \dfrac{\nu}{3}\x + \left(\dfrac{2\nu^2}{9}-\dfrac{\nu}{4}\right)\x^2
   + \left(\dfrac{14}{81}\nu^3+\dfrac{\nu^2}{3}-\dfrac{9\nu}{8}\right)\hat{\hat{x}}^3 \bigg],
\end{align}
that is obtained by inverting the 3PN-accurate expansion of Eq.~\eqref{eq:xbar_of_x}, and, as a last
step, we crucially {\it factorize} out from the resulting modified Eq.~\eqref{eq:Finfty} the Newtonian
prefactor considered now as an {\it explicit} function of $\x$,
$F_\lm^{(N,\epsilon)}(\x)\propto \x^{\ell+3+\epsilon}$. It follows that a new
residual amplitude corrections $\tilde{\rho}_\lm^\infty(\x;\,\nu)$ is defined 
as the following expansion in powers of $\x$
\be
\label{eq:def_rho_tilde}
\tilde{\rho}_\lm^\infty(\x;\, \nu) = {\rm Taylor}_N\left\{ f(\x)^{(\ell+3+\epsilon)/(2\ell)}\rho_\lm^\infty[\bar{x}_{\rm 3PN}(\x)]\right\},
\ee
where the notation ${\rm Taylor}_N\left\{...\right\}$ indicates the Taylor expansion of the argument inside
the curly brackets at order $N$. Here $N$ is the highest power present in the
Taylor-expanded $\rho_\lm^\infty(x;\,\nu)$ in Eq.~\eqref{eq:hatF}.
At the end, a new, different, $\x$-dependent, expression for the asymptotic
multipolar flux that is well-behaved along both stable and unstable circular orbits is
\begin{align}
\label{eq:Finfty_unstable}
\tilde{F}^{(\infty,\epsilon)}_\lm(\x;\,\nu) & =  F_\lm^{(N,\epsilon)}(\x;\,\nu)\nonumber\\
                     &\times \left(\hat{S}_{\rm eff}^{(\epsilon)}(x(\x))\right)^2|T_\lm(\x)|^2\left[\tilde{\rho}_\lm^\infty(\x;\, \nu)\right]^{2\ell}.
\end{align}
Note that the $\nu$-dependent coefficients of the Taylor-expanded function $\tilde{\rho}_\lm^\infty (\x;\,\nu)$
are different from the corresponding ones of the $\rho_\lm^\infty (x;\,\nu)$. On the contrary,
in the test-mass limit (where $\x=x$) one has  $\tilde{\rho}_\lm^\infty (\x;\,0)= \rho_\lm^\infty (x;\,0)$
(and thus $\tilde{F}^{(\infty,\epsilon)}_\lm(\x;\,0)=F^{(\infty,\epsilon)}_\lm(\x;\,0)$).
Since we shall need it below, we explicitly write the expression
for $\tilde{\rho}_{22}^\infty(\x;\,\nu)$ expanded formally to 3PN (in the variable $\x$) accuracy
\begin{widetext}
\begin{align}
\label{eq:tilde_rho}
\tilde{\rho}_{22}^\infty(\x;\,\nu)=1 & + \x \left(\frac{15}{14}\nu-\frac{43}{42}\right)
+\x^2 \left(\frac{13217}{10584}\nu^2-\frac{27947
   }{10584}\nu-\frac{20555}{10584}\right)\nonumber\\
& +\x^3 \left(\frac{2347187}{1629936}\nu^3-\frac{10439395}{2444904}\nu^2+\frac{41 \pi ^2 \nu }{192}
-\frac{38845027}{4889808}\nu-\dfrac{428}{105}{\rm eulerlog}_2(\x)+\frac{1556919113}{122245200}\right)+\O(\x^4),
\end{align}
\end{widetext}
where ${\rm eulerlog}_m(\x) = \gamma + \log 2 +\dfrac{1}{2}\log \x + \log m$.
Evidently Eqs.~\eqref{eq:def_rho_tilde}-\eqref{eq:tilde_rho} define {\it only one}
among the many possible Taylor-expanded representations of
$\tilde{\rho}_{22}^{\infty}(\x;\,\nu)$. For example, to quote one alternative
possibility,  we could have used the Taylor-expansion
of Eq.~\eqref{eq:xbar_of_x} to express $\x$ as a function of $x$ and then use again
the non-expanded implicit relation $x(\x)$ to represent the result as a function
of $\x$. Whatever one does for representing the relativistic correction, the crucial
new element that makes the flux well behaved also below the LSO is the factorization of
the Newtonian contribution as an explicit function of $\x$ and {\it not} of $x$.
Note that, for simplicity, we did not include in Eq.~\eqref{eq:tilde_rho} also the
4PN and 5PN test-mass terms that can be used to compute the hybrid $3^{+2}$PN
approximation to $\tilde{\rho}_{22}^\infty(\x;\,\nu)$. They can be found in Eq.~(50)
of Ref.~\cite{Damour:2008gu} and added to Eq.~\eqref{eq:tilde_rho}
as $\x^4$ and $\x^5$ terms.

\subsection{Factorization of the multipolar horizon flux}
\label{sec:rho_lm_H}

We have seen in the previous section that after factorization all information
about the asymptotic GW amplitude and asymptotic GW energy flux is contained in the
$\rho_\lm^\infty$ (or $\tilde{\rho}_\lm^\infty$) functions that are therefore the
only real unknowns of the problem (for a given factorization).
Let us focus now on the horizon-absorbed flux.
The strategy that we shall follow here is to introduce a resummed
factorization procedure for the multipolar horizon fluxes that is analogous to
the one used above for the asymptotic fluxes. The main idea is to
factorize out from the PN-expanded representation of the horizon fluxes,
given by Eqs.~\eqref{eq:PS94},\eqref{eq:F22Hpn} and~\eqref{eq:F21Hpn},
both the leading-order contributions (analogous to the Newtonian prefactor
in the asymptotic fluxes) and the effective source so to eventually introduce
some residual (horizon) amplitude corrections $\rho_\lm^H(x;\,\nu)$ of the form
$1+x+x^2+\dots$ (or correspondingly $\tilde{\rho}_\lm^H(\x;\,\nu)=1+\x+\x^2+\dots$)
that take into account the remaining PN corrections.
As in the previous section, we first present a factorization
that is well-behaved for stable circular orbits only. We then 
start from this result to obtain a different  factorization of the
horizon flux where the leading-order prefactor is an explicit function
of $\x$ and so it is meaningfully defined for stable and unstable circular orbits.

We write each multipolar contribution to the horizon flux as
\be
F_{\ell m}^{(H,\epsilon)}(x;\,\nu) = F_\lm^{(H_{\rm LO},\epsilon)}(x;\,\nu) \F^{(H,\epsilon)}_\lm(x;\,\nu),
\ee
where $F_\lm^{(H_{\rm LO},\epsilon)}(x;\,\nu)$ indicates the leading order contribution to
the absorbed fluxes (given by Eqs.~\eqref{eq:F22LO}-\eqref{eq:F21LO} for $\ell=2$
and~Eq.~\eqref{eq:PS94} for the other multipoles\footnote{Note that in this
context $\nu$ is always assumed to be the symmetric mass ratio 
 $\nu=m_A m_B/M^2$, even if Eq.~\eqref{eq:PS94} is obtained in the
test-particle limit. In doing so we are in fact neglecting the
(yet unknown) polynomials in $\nu$, of the form $1+\nu + \nu^2 +\dots$,
that have to appear in the LO prefactors when the $\nu$-dependent
calculation of Eqs.~\eqref{eq:F22LO}-\eqref{eq:F21LO} is extended
to the other multipoles.}) and $\hat{F}^{(H,\epsilon)}_\lm(x;\,\nu)$
accounts for all higher-order relativistic corrections.
We then factorize $\hat{F}^{(H,\epsilon)}_\lm(x;\,\nu)$ as
\be
\label{eq:Fhlm}
\hat{F}^{(H,\epsilon)}_\lm(x;\,\nu) = \left(\hat{S}_{\rm eff}^{(\epsilon)}(x)\right)^2
\left(\rho_\lm^H(x;\,\nu)\right)^{2\ell}.
\ee
The physical reason for factorizing the effective source in $\hat{F}^{(H,\epsilon)}_\lm(x;\,\nu)$
is the same one done in Ref.~\cite{Damour:2008gu} for
$\hat{F}^{(\infty,\epsilon)}_\lm(x;\,\nu)$. The idea in Ref.~\cite{Damour:2008gu}
was motivated  by the form of the equation satisfied by each partial wave in the
circular test-mass limit with a source made up of the important
dynamical constituents of (test-particle) dynamics,
i.e. its energy and angular momentum (see discussion below Eq.~(8) of~\cite{Damour:2008gu}).
Note that, as it is the case with $\hat{F}^{(\infty,\epsilon)}_\lm(x;\,\nu)$, the factorization of
the source for each multipole actually means factorizing out the singular behavior of the
absorbed energy flux at the (EOB-defined) light-ring\footnote{This idea is similar to the suggestion of
Ref.~\cite{Damour:1997ub} of factoring out a pole from the energy flux.}.
Notably, Ref.~\cite{Bernuzzi:2011aj} has shown that, in the test-mass limit,
the numerical $\rho_\lm^\infty(x;0)$ functions appear
to be regular when $x$ is close to $1/3$. This a posteriori confirms the intuition
that the singular behavior of the flux at the light-ring is only due to the squared source.
Similarly, in proposing here the factorization~\eqref{eq:Fhlm}, we are assuming
the same singular behavior also for the horizon fluxes, so that we expect the functions
$\rho_\lm^H(x;\,0)$ to be regular when $x$ is close to $1/3$. In Sec.~\ref{sec:test_mass} we
will see that this intuition is supported by the behavior of the ``exact'' test-mass $\rho_\lm^H(x;\,0)$
computed numerically\footnote{In this respect, note that the absorbed flux in the small-hole
approximation computed by Poisson in Ref.~\cite{Poisson:2004cw}, Eq.~(8.52), has a rather
different structure since the singular behavior is proportional to the {\it fourth power}
of the source.}.

Let us now compute the explicit analytical expressions of the  $\rho_\lm^H(x;\,\nu)$'s.
Evidently, since (1PN) fractional corrections have been analytically computed only
for the $\ell=m=2$ PN-expanded horizon flux, Eq.~\eqref{eq:F22Hpn},
one can obtain them only for $\rho_{22}^H(x;\nu)$ and only at 1PN fractional accuracy,
while $\rho_\lm^H(x;\,\nu)=1$ (i.e., 0PN fractional accuracy) for all other multipoles.
By factoring  out of Eq.~\eqref{eq:F22Hpn} the effective source, $\hat{H}_{\rm eff}(x;\,\nu)$,
and the leading-order contribution, $F_{22}^{H_{\rm LO}}(x;\,\nu)=(32/5)\nu^2(1-4\nu+2\nu^2)x^9$,
taking the fourth root and then expanding this latter at 1PN accuracy, we obtain
\be
\label{eq:rho22_x}
\rho_{22}^H(x;\,\nu) = 1+\dfrac{4-21\nu+27\nu^2-8\nu^3}{4(1-4\nu+2\nu^2)}x+\O(x^2).
\ee
The combination of Eq.~\eqref{eq:Fhlm} and Eq.~\eqref{eq:rho22_x} gives
an analytical expression of the multipolar horizon flux that, as it was the case for
Eq.~\eqref{eq:flux_std} is ill-behaved along unstable orbits. We need then
to apply the same procedure that lead to Eq.~\eqref{eq:Finfty_unstable}
for the asymptotic flux. We first make explicit the dependence on $\x$ in the
horizon flux
\begin{align}
F_\lm^{(H,\epsilon)}(x(\x);\,\nu)&= F_\lm^{(H_{\rm LO},\epsilon)}(x(\x);\,\nu)\nonumber\\
                            &\times \left(\hat{S}_{\rm eff}^{(\epsilon)}(x(\x))\right)^2
\left(\rho_\lm^H(x(\x);\,\nu)\right)^{2\ell},
\end{align}
then we replace in both  $F_\lm^{(H_{\rm LO},\epsilon)}(x(\x);\,\nu)$ and $\rho_\lm^H(x(\x);\,\nu)$ the function
$x(\x)$ with $\bar{x}_{\rm 3PN}(\x)$, Eq.~\eqref{eq:xofxhat}, and in the end we factorize in front
$F^{(H_{\rm LO},\epsilon)}_\lm(\x)$ as an explicit function of $\x$. Since this factor is proportional
to $\x^{2\ell+5+\epsilon}$, this operation yields a new residual amplitude correction
as the following expansion in powers of $\x$
\be
\tilde{\rho}^H_\lm(\x;\,\nu)={\rm Taylor}_N\left\{f(\x)^{(2\ell+5+\epsilon)/(2\ell)}\rho_\lm^H[\bar{x}_{\rm 3PN}(\x)]\right\},
\ee
where $N$ is now the highest power present in the Taylor-expanded
$\rho_\lm^H(x;\,\nu)$ of Eq.~\eqref{eq:Fhlm} above. Eventually the new, $\x$-dependent,
expression for the horizon multipolar flux that is well-behaved along stable and
unstable orbits is
\begin{align}
\label{eq:FHunstable}
\tilde{F}^{(H,\epsilon)}_\lm(\x;\,\nu) & =  F_\lm^{(H_{\rm LO},\epsilon)}(\x;\,\nu)\nonumber\\
                           &\times\left(\hat{S}_{\rm eff}^{(\epsilon)}(x(\x))\right)^2\left[\tilde{\rho}_\lm^H(\x;\, \nu)\right]^{2\ell}.
\end{align}
In the $\ell=m=2$ case we have
\be
\label{eq:rho22_anlyt}
\tilde{\rho}_{22}(\x;\,\nu) = 1 + \dfrac{4-18\nu+15\nu^2 - 2\nu^3}{4(1-4\nu+2\nu^2)}\x + \O(\x^2),
\ee
while it is $\tilde{\rho}_\lm(\x;\,\nu)=1$ for all other multipoles.
For $\nu=0$, where $\x=x$ and thus $\rho_\lm^H(x;\,0)=\tilde{\rho}_\lm^H(x;\,0)$),
this equation reduces to the very simple result $\rho_{22}^H(x;\,0)=1+x+\O(x^2)$.
Note also that the dependence on $\nu$ is relatively mild, since the coefficient
of $\x$ in Eq.~\eqref{eq:rho22_anlyt} varies between 1,
when $\nu=0$, and $13/16\approx 0.8125$, when $\nu=1/4$.
This corresponds to less than a $5\%$ fractional difference at the test-mass
light-ring position, $\x=1/3$ between $\nu=0$ and $\nu=1/4$.
In the next Section we shall check the consistency of the 1PN analytical prediction
and its accuracy with respect to the $\rho_{22}^H(x;\,0)$ computed numerically.

\section{Numerical computation of $\rho_\lm^H(x;\,0)$.}
\label{sec:test_mass}
\begin{figure}[t]
\center
\includegraphics[width=0.45\textwidth]{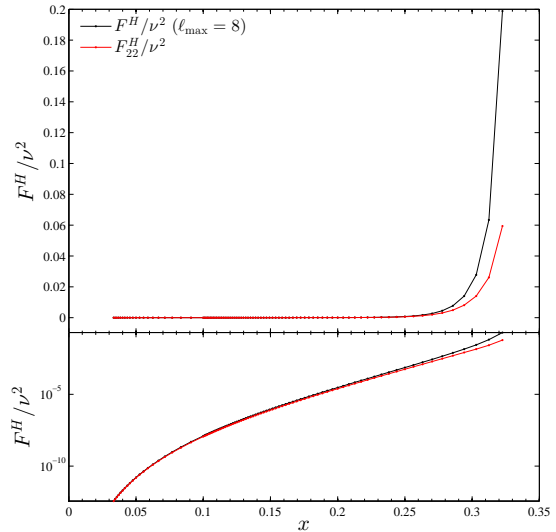}
\caption{
\label{fig:FH_0}Horizon-absorbed energy flux for a point-particle orbiting along stable and unstable
circular orbits of the Schwarzschild spacetime. The total flux (summed up to $\ell_{\rm max}=8$) is
contrasted with the quadrupolar $\ell=m=2$ flux. The same quantities are plotted on a linear
(top) and logarithmic (bottom) vertical scale. Here $x=1/r$, where $r$ is the Schwarzschild radial
coordinate. The light-ring is located at $x=1/3$.}
\end{figure}
\begin{figure*}[t]
\center
\includegraphics[width=0.45\textwidth]{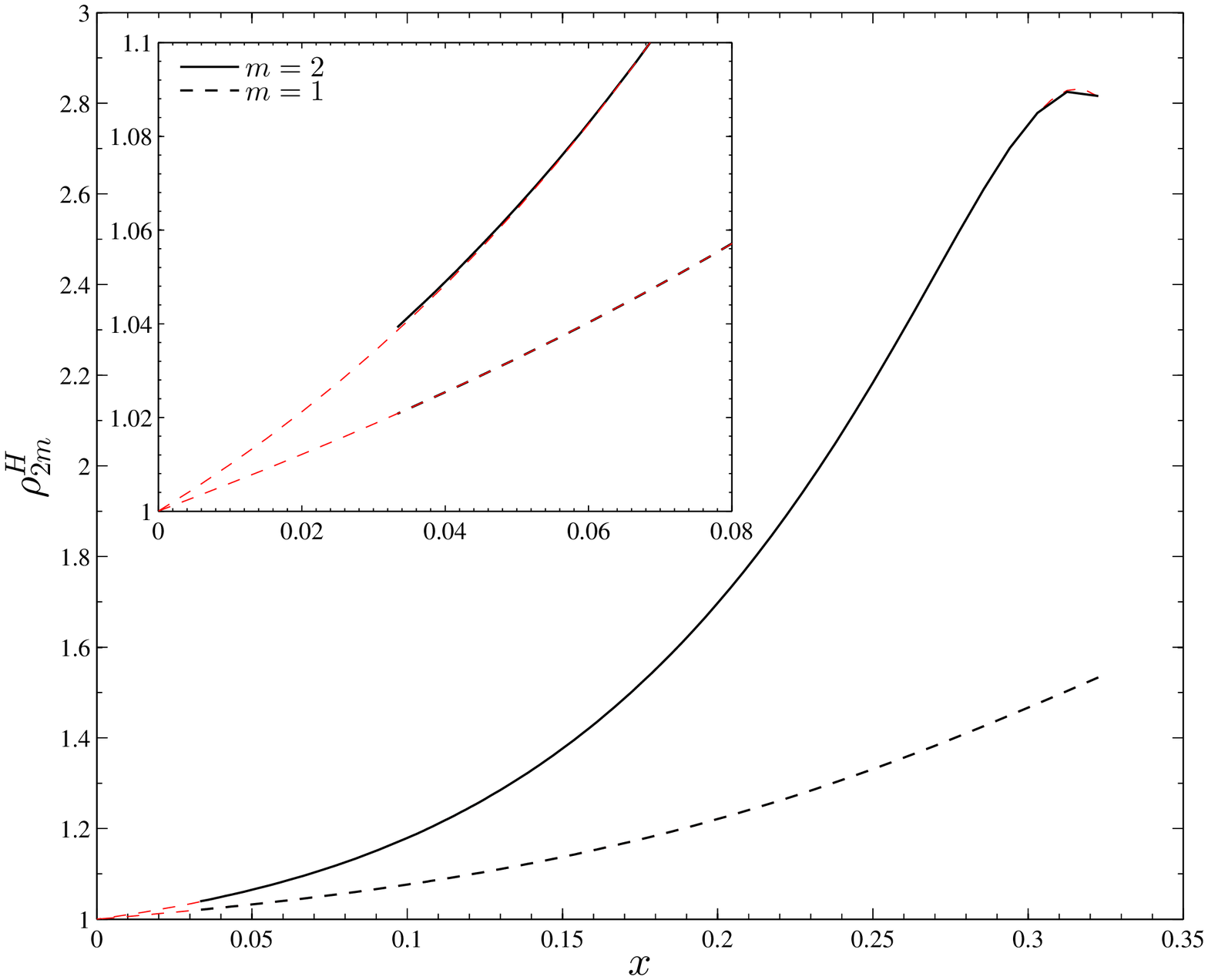}
\includegraphics[width=0.45\textwidth]{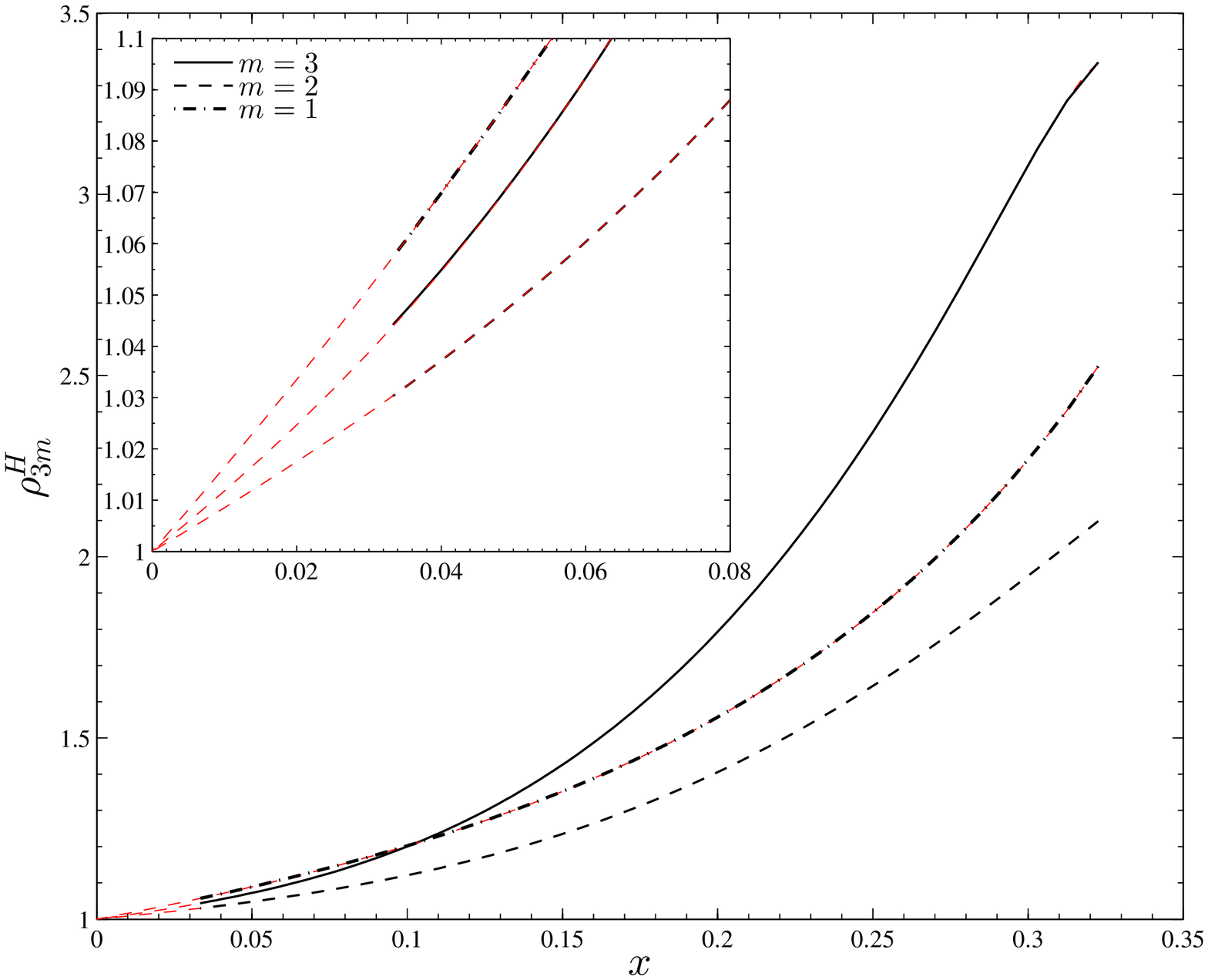}\\
\includegraphics[width=0.45\textwidth]{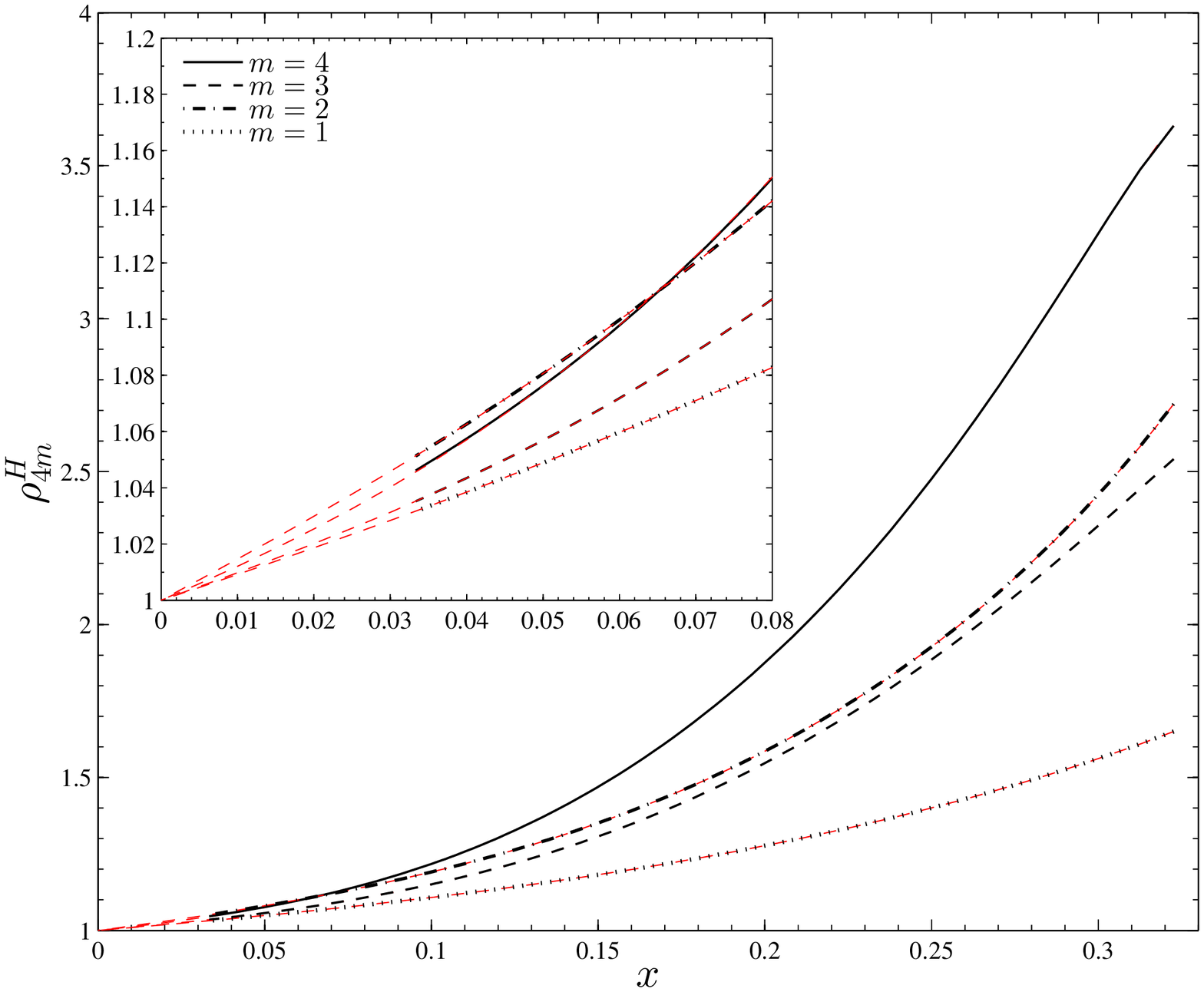}
\includegraphics[width=0.45\textwidth]{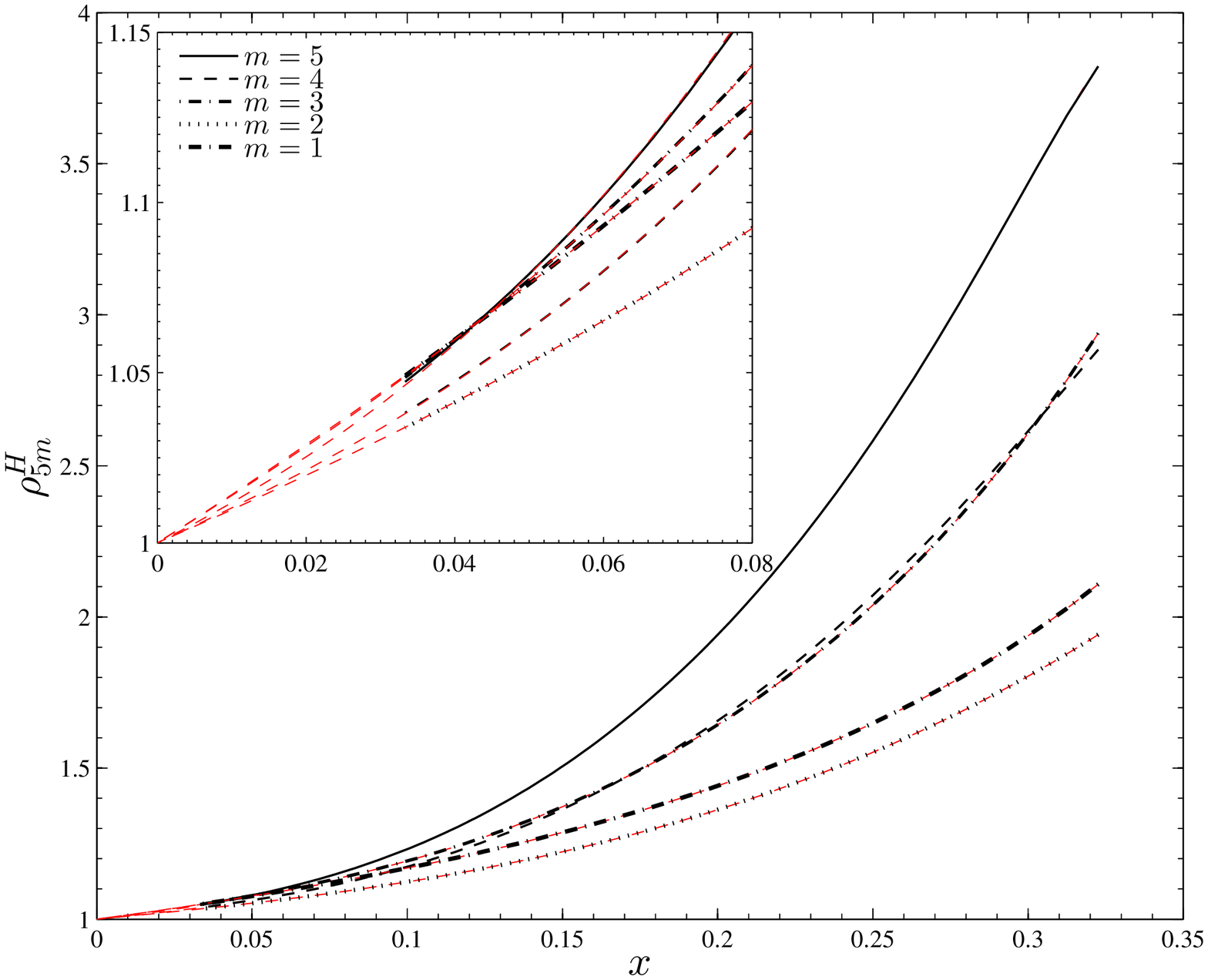}\\
\includegraphics[width=0.45\textwidth]{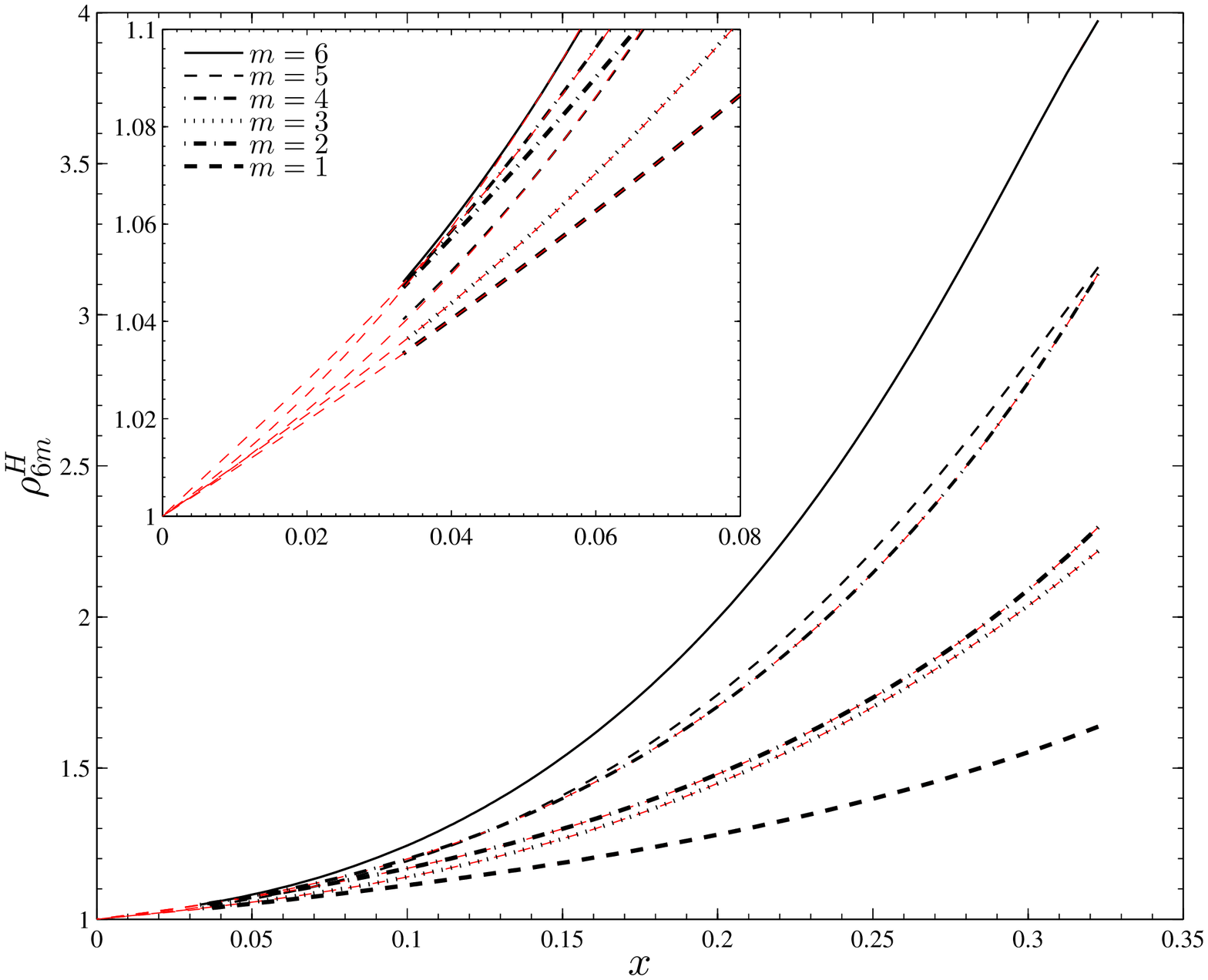}
\includegraphics[width=0.45\textwidth]{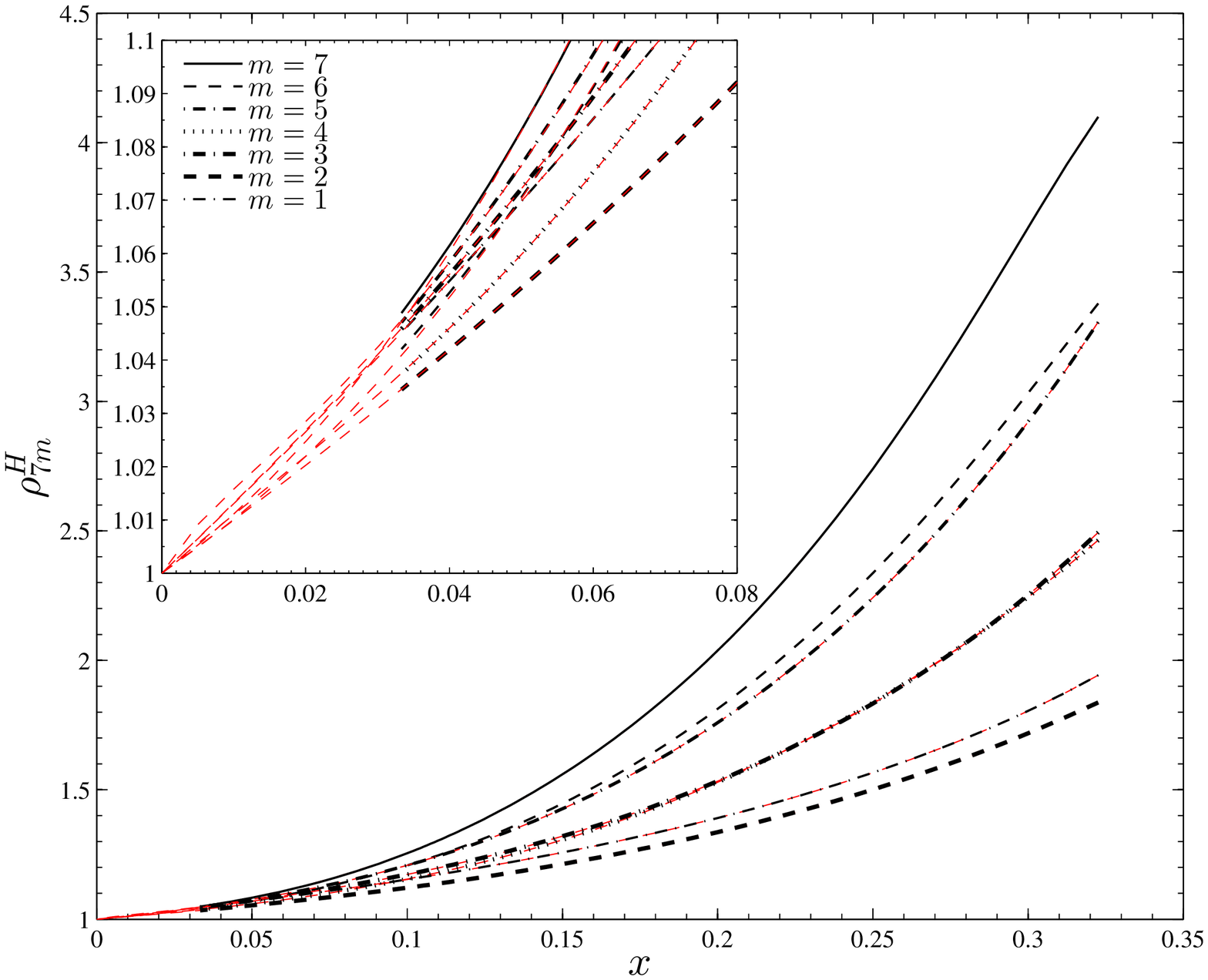}
\caption{The numerical $\rho_\lm^H(x;\,0)$ computed from the numerical fluxes using Eq.~\eqref{eq:rho_num}.
The dashed (red online) lines indicate the outcome of the fit using the rational function given in
Eq.~\eqref{eq:rhofit}.
\label{fig:rholmH}}
\end{figure*}
\begin{table*}[t]
  \caption{\label{tab:coeff} Values of the coefficients of the fitting function Eq.~\eqref{eq:rhofit} for
multipoles up to $\ell=7$.}
\begin{center}
\begin{ruledtabular}
\begin{tabular}{cccccccc}
$\ell$ &  $m$ & $n_1^{\lm}$ & $n_2^{\lm}$ &  $n_3^{\lm}$  &  $n_4^{\lm}$  & $d_1^{\lm}$  & $d_2^{\lm}$ \\
\hline
2 &2  & -3.78016&     7.02291 &   0.70963 & -33.06646 &  -4.71422 &   5.60892\\
2 &1  & -1.89997&     0.49519 &   5.98688 & -17.15286 &  -2.48118 &   0.92669\\
3 &3  & -2.36556&     1.73143 &  34.37095 &-124.89266 &  -3.50205 &   1.87045\\
3 &2  & -1.47910&    -0.37517 &  20.54194 & -56.00960 &  -2.31621 &   0.16676\\
3 &1  & -1.50854&     0.24240 &   4.47148 &  -9.36740 &  -3.11918 &   2.29453\\
4 &4  & -1.96477&     1.77455 &  41.82985 &-155.50256 &  -3.11767 &   0.77264\\
4 &3  & -0.88873&    -2.35228 &  38.83982 &-118.59113 &  -1.84937 &  -2.03044\\
4 &2  & -0.35892&    -0.14473 &  17.55970 & -28.91807 &  -1.79350 &  -0.00413\\
4 &1  & -1.86354&     0.16555 &   3.23144 &  -8.41344 &  -2.76941 &   1.49953\\
5 &5  & -2.01738&     3.91457 &  35.23863 &-150.27033 &  -3.12766 &   0.87106\\
5 &4  & -0.89483&    -1.79998 &  44.98785 &-154.35735 &  -1.88958 &  -2.54359\\
5 &3  &  3.96179&    -1.42399 &  50.85186 & -23.44662 &   2.58708 &  -5.91327\\
5 &2  & -0.79955&    -1.18556 &  13.74389 & -34.03313 &  -1.76133 &  -0.62741\\
5 &1  & -1.68170&     0.13047 &   2.74892 &  -6.05858 &  -3.07511 &   2.19172\\
6 &6  & -2.27965&     6.83094 &  21.35950 &-125.03545 &  -3.33171 &   1.53708\\
6 &5  & -1.35000&     1.11200 &  36.38329 &-148.13578 &  -2.32261 &  -1.50432\\
6 &4  & 17.75737&    -4.20987 & 160.53963 & 101.12617 &  16.32248 & -21.05465\\
6 &3  &  4.39780&     1.01247 &  43.62830 &   2.92466 &   3.37649 &  -2.66169\\
6 &2  & -0.52866&    -0.68194 &  11.73839 & -24.81141 &  -1.85746 &  -0.15371\\
6 &1  & -1.88052&     0.17089 &   2.21690 &  -5.55880 &  -2.83756 &   1.68356\\
7 &7  & -2.60076&     9.77862 &   5.23594 & -92.25665 &  -3.59794 &   2.37691\\
7 &6&   -1.90285&     4.69608 &  20.96354 &-120.14340 &  -2.83491 &  -0.03207\\
7 &5&  159.88998&   -25.98598 &1348.46529 &1881.03562 & 157.35115 &-161.73362\\
7 &4&   19.50198&    10.68544 & 143.28131 & 252.11345 &  18.36505 &  -4.37623\\
7 &3&    2.53654&    -2.61813 &  33.54445 & -50.19582 &   1.23298 &  -5.23258\\
7 &2&   -0.71568&    -1.37276 &   9.79925 & -25.74536 &  -1.70002 &  -0.80200\\
7 &1&   -1.76491&     0.15087 &   1.99348 &  -4.43712 &  -3.04998 &   2.14649\\
\end{tabular}
\end{ruledtabular}
\end{center}
\end{table*}
\begin{table}[t]
  \caption{\label{tab:c1PN}Numerical evaluation of the 1PN coefficient
$c_1^\lm$ in Eq.~\eqref{eq:rhofit} from the fit to numerical data. 
Note that $c_1^{22}$ differs from its analytically predicted value 
(i.e., $c_1^{22}=1$ as obtained from the $\nu=0$ specification of 
Eq.~\eqref{eq:rho22_anlyt}), by a $6.6\%$ due to the global nature 
of the fit.}
\begin{center}
\begin{ruledtabular}
\begin{tabular}{ccccc}
&$\ell$ &  $m$ & $c_1^{\lm}$ \\
\hline
&2 & 1 &    0.58121 \\
&2 & 2 &    0.93406\\
&3 & 1 &    1.61064\\
&3 & 2 &    0.83711\\
&3 & 3 &    1.13649\\
&4 & 1 &    0.90588\\
&4 & 2 &    1.43458\\
&4 & 3 &    0.96063\\
&4 & 4 &    1.15290\\
&5 & 1 &    1.39341\\
&5 & 2 &    0.96178\\
&5 & 3 &    1.37472\\
&5 & 4 &    0.99476\\
&5 & 5 &    1.11029\\
&6 & 1 &    0.95704\\
&6 & 2 &    1.32879\\
&6 & 3 &    1.02131\\
&6 & 4 &    1.43489\\
&6 & 5 &    0.97261\\
&6 & 6 &    1.05206\\
&7 & 1 &    1.28508\\
&7 & 2 &    0.98434\\
&7 & 3 &    1.30356\\
&7 & 4 &    1.13693\\
&7 & 5 &    2.53883\\
&7 & 6 &    0.93206\\
&7 & 7 &    0.99718\\
\end{tabular}
\end{ruledtabular}
\end{center}
\end{table}

Since the highest PN analytical information available for the $\tilde{\rho}_\lm^H$'s
is given by the 1PN-accurate result of Eq.~\eqref{eq:rho22_anlyt}, that is
limited to the $\ell=m=2$ multipole,
the first question that one wants to address is about its accuracy
in the strong-field, fast-velocity regime. In the test-mass limit
this question can be answered  exhaustively by comparing  Eq.~\eqref{eq:rho22_anlyt}
with the numerical $\rho_{22}^H(x;\,0)$
that can be extracted from the numerical computation of the GW absorbed flux,
in a way similar to what was done in Ref.~\cite{Damour:2008gu} for the asymptotic fluxes.
To this aim, we computed the horizon absorbed energy flux for a selected
sample of circular orbits belonging to both the stable and unstable branch.
Although the horizon flux has been computed in the past for stable circular
orbits~\cite{Finn:2000sy,Martel:2003jj,Barack:2007tm}, we are not aware of any published results
for unstable orbits. We computed the horizon fluxes using the gravitational self-force (GSF) code developed in
Ref.~\cite{Akcay:2010dx}, which is a frequency domain variant of the original
self-force code of Ref.~\cite{Barack:2007tm}. The GSF code used computes
the $\mathcal{O}(\mu/M)$ metric perturbation. The flux is then obtained in terms
of surface integrals of Weyl scalars evaluated at the event horizon of the large mass.
For circular orbits, the expressions for the horizon and asymptotic fluxes go back
to Teukolsky's work in the 1970s~\cite{Teukolsky:1973, Teukolsky:1974}.
We consider orbits with radii varying from $r_{\rm min}=3.1$ to $r_{\rm max}=30$,
that are spaced by $\Delta r=0.1$ for  $r\in[3.1,10]$ and by $\Delta r=1$
for $r\in[10,30]$. We have computed the numerical $F_\lm^H/\nu^2$ up to $\ell_{\rm max}=8$:
Fig.~\ref{fig:FH_0} compares the total flux (summed up to $\ell_{\rm max}=8$)
with the $\ell=m=2$ contribution. The frequency domain approach allows us to obtain
the $F_\lm^{H_{\rm num}}/\nu^2$'s very accurately. In particular, the fractional uncertainty
is extremely low, i.e. on the order of $10^{-10}$  or smaller for all multipoles and
for strong-field orbits (say $r\leq 10$); it is at most of order $10^{-2}$-$10^{-3}$ for
the $l=8$, $m=(2,3)$ multipoles when $r=30$. This apparent ``inaccuracy'' is due to
the fact that beyond $r\sim 15$, the magnitude of $(\ell,m)$ modes of horizon flux is
comparable to or less than machine accuracy. So the concept of ``fractional accuracy''
becomes numerically meaningless. For example, the $\ell=8$, $m=1$ multipolar contribution
is ($\sim 10^{-33}$). 

The $\rho_\lm^H(x;\,0)$ are computed from
the horizon numerical fluxes $F_\lm^{H_{\rm num}}$ 
simply by
\be
\label{eq:rho_num}
\rho_\lm^H(x;\,0) = \dfrac{\sqrt{F_\lm^{H_{\rm Num}}/F_\lm^{(H_{\rm LO},\epsilon)}(x;\,0)}}{ \hat{S}_{\rm eff}^{(\epsilon)}(x) },
\ee
where in the test-mass limit one has
\begin{align}
\hat{S}^{(0)}_{\rm eff}(x)=\dfrac{1-2 x}{\sqrt{1-3x}},\\
\hat{S}^{(1)}_{\rm eff}(x)=\dfrac{1}{\sqrt{1-3x}}.
\end{align}
The result of this computation is shown in Fig.~\ref{fig:rholmH} for multipoles up to $\ell=7$.
An important feature highlighted by the plot is that the value of $\rho_\lm^H$ increases by a
considerable factor (between 3 and 4) when $x$ varies from $1/30$ to $1/3.1$. Such a behavior
is rather different from the quasi-linear (decreasing) trend of the numerical $\rho_\lm^\infty$ found in
Refs.~\cite{Damour:2008gu,Bernuzzi:2011aj} (see Fig.~3 and~4 of~\cite{Damour:2008gu} and
Fig.~7 of~\cite{Bernuzzi:2011aj}).
This fact shows that the 1PN approximation of $\rho_{22}^H$ will be rather rough and
more analytical information is needed to reproduce  the correct behavior of the $\rho_\lm^H(x;\,0)$
of Fig.~\ref{fig:rholmH} towards the light-ring (see also the discussion
around Fig.~\ref{fig:rho22H} below).

In the absence of high-order PN results for the horizon fluxes that allowed us to push
the analytical knowledge of $\rho_\lm^H$ to higher order, for the moment we content
ourselves to have an effective analytical representation of the numerical results
in terms of simple functions. A convenient procedure to do so is to fit the numerical
data with a suitable function that tends to 1 when $x\to 0$.
We found that the following rational function
\be
\label{eq:rhofit}
\rho_\lm^H(x) = \dfrac{1+n_1^\lm x + n_2^\lm x^2 + n_3^\lm x^3 + n_4^\lm x^4}{1 + d_1^\lm x + d_2^\lm x^2}
\ee
works well for all multipoles (and especially for the $\ell=m=2$ one, which presents a characteristic ``bump''
close to the light ring, similarly to the $\rho_\lm(x;\,0)$'s~\cite{Bernuzzi:2011aj}).
The coefficients are listed in Table~\ref{tab:coeff} for multipoles up to $\ell=7$.
One notes that their values are (on average) relatively small due to the choice of a rational function
to perform the fit. Had we used instead a polynomial expression (with the same number of free parameters)
we would have obtained much larger numerical values.

From Eq.~\eqref{eq:rho22_anlyt} we have $\rho_{22}^H(x;\,0)=1+ x + \O(x^2)$.
It is then interesting to compute a PN approximation of the fit to
$\rho_{22}^H$, in order to have a concrete idea of the reliability of
the ansatz~\eqref{eq:rhofit}. By expanding Eq.~\eqref{eq:rhofit}
around $x=0$ we find
\be
\rho_{\lm}^H(x) = 1 + c_1^{\lm} x + c_2^{\lm}x^2 +c_3^{\lm}x^3 + c_4^{\lm}x^4+ \O(x^5),
\ee
where $c_1^{\lm}=n_1^\lm-d_1^\lm $, $c_2^{\lm}=(d_1^\lm)^2-d_2^\lm - d_1^\lm n_1^\lm + n_2^\lm$
and the corresponding dependence of the other $c_i^\lm$ on $n_i^\lm$ and $d_i^\lm$.
The numerical values of $c_{1}^{\lm}$ are listed in Table~\ref{tab:c1PN} for completeness.
First of all, one immediately notices that the $c_1^\lm$ are always of order unity.
The smallest value is $c_1^{21}=0.58121$, and the largest $c_1^{75}=2.53883$,
where the fit does not seem to be visually very accurate
when $x\to 0$. Inspection of Table~\ref{tab:c1PN} shows that the value 1 is a
good approximation for several (actually, almost all) multipoles.
In particular, one has $c_1^{22}=0.93406$, which is smaller only by a $6.6\%$
with respect to the analytically predicted value.
This, rather small discrepancy is due to the global nature of the fit.
Note in fact that this agreement can be further improved by including also a $x^5$--type
term in the numerator of Eq.~\eqref{eq:rhofit}, so to allow for more flexibility.
In this case, the coefficients of the fit take the following numerical values
$(n_1,n_2,n_3,n_4,n_5,d_1,d_2)=(-4.59,6.01,26.95,-178.53 ,276.68,-5.60,8.05)$, which yield
$c_1^{22}=1.01590$. However, the fact that the $n_4$ and $n_5$ coefficients become
rather large (and relatively close) is probably an indication that we are allowing
for excessive freedom in the analytical model.
For this reason we prefer to work with a 4th-order polynomial at the numerator
even if the results is slightly less accurate when compared to the 1PN analytical value.

An alternative possibility is to {\it impose} the analytical 1PN value as a constraint
in the fit, i.e. by asking that $d_1^{22}=n_1^{22}-1$. In doing so, the fit improves
(though at a level indistinguishable on the scale of Fig.~\ref{fig:rholmH},
see Fig.~\ref{fig:rho22H}) and the low-frequency limit is recovered correctly
by construction. In this case one gets the following numerical values
$(n_1,n_2,n_3,n_4,n_5,d_2)=(-3.51327,5.28819,10.16670,-52.90979,5.01393)$.
\begin{figure}[t]
\center
\includegraphics[width=0.5\textwidth]{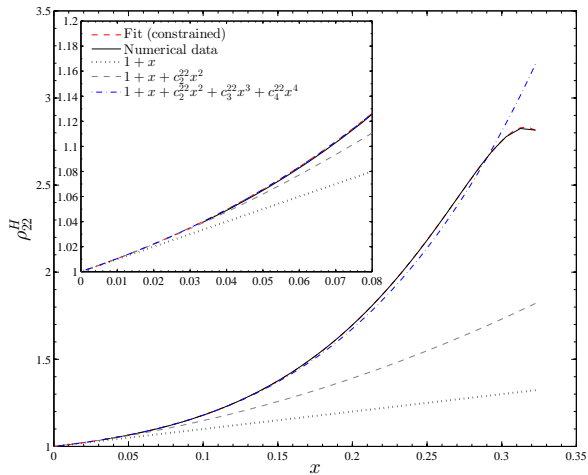}
\caption{Test-mass limit: investigating the properties of $\rho_{22}^H(x;\,0)$.
The ``exact'' numerical result is contrasted with: (i) the 1PN-accurate analytical
prediction; (ii)  the fit of the numerical data by the rational function given by
Eq.~\eqref{eq:rhofit} with the 1PN-constraint $d^{22}_1 = n_1^{22}-1$;
(iii) the 2PN-accurate and 4PN-accurate expansions of this fitting function.
\label{fig:rho22H}}
\end{figure}
\begin{figure*}[t]
\center
\includegraphics[width=0.45\textwidth]{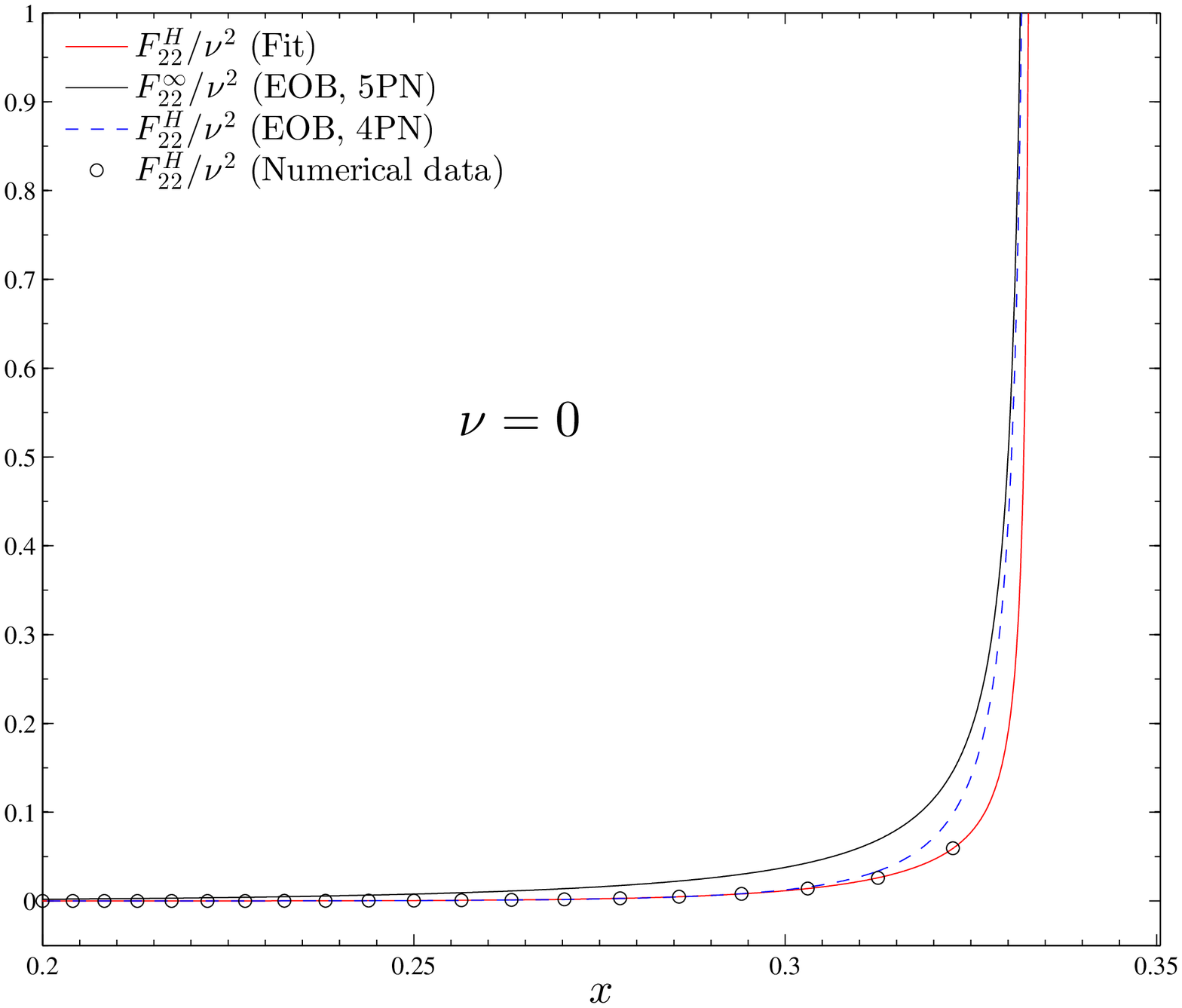}
\includegraphics[width=0.45\textwidth]{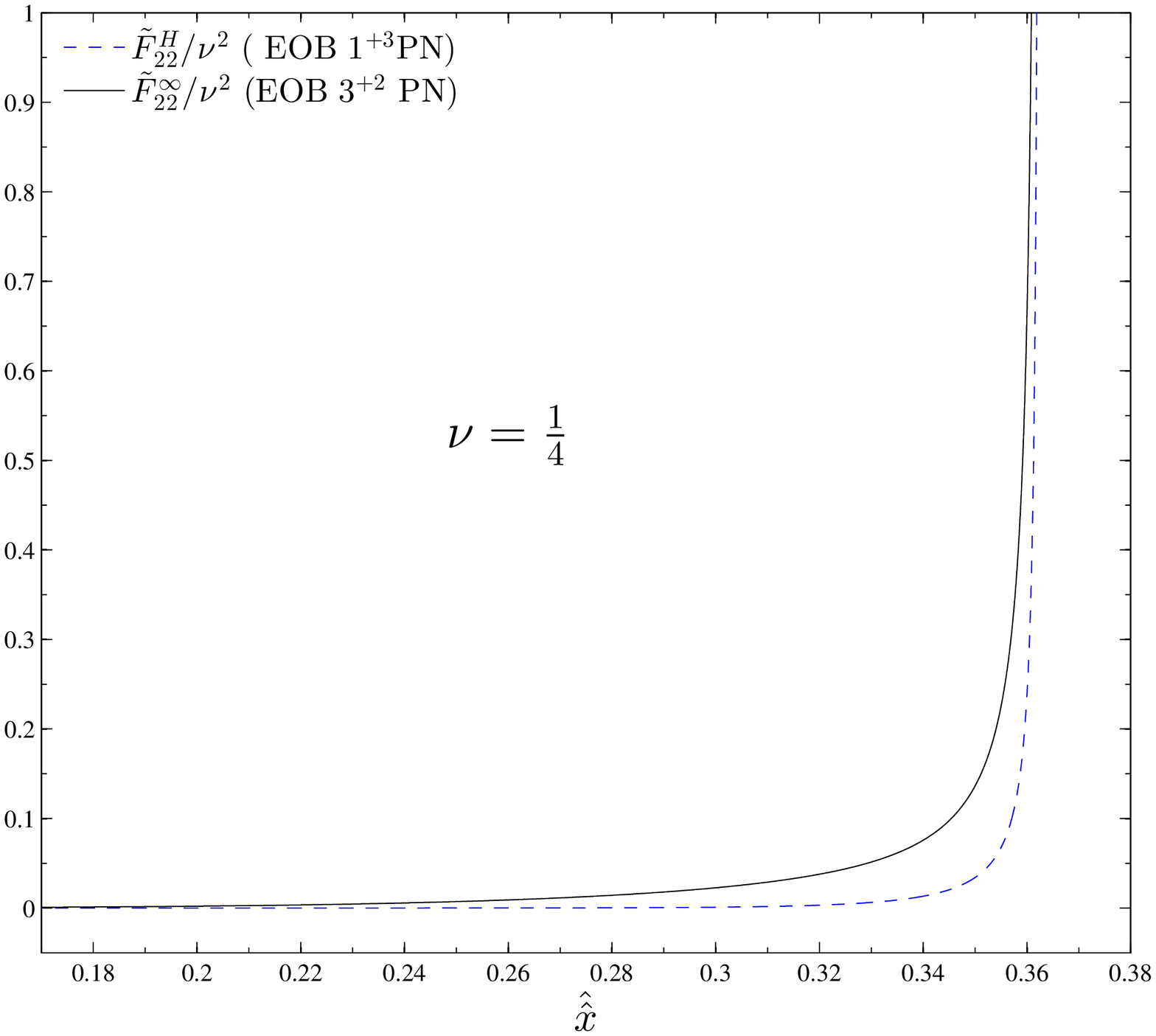}
\caption{Comparison between asymptotic and horizon fluxes (divided by $\nu^2$)
computed in the EOB approach. The left panel refers to the test-mass limit, while the right-panel
to the equal-mass case. The left panel also highlights the difference in the horizon flux between
the fit representation of $\rho^H_{22}(x;\,0)$ and its 4PN-accurate expansion.
See text for further explanations.
\label{fig:fluxes}}
\end{figure*}
Figure~\ref{fig:rho22H} focuses on the $\ell=m=2$ mode only and in particular it illustrates
the performance of such a constrained fit.
The exact data (black solid line) are contrasted with the analytical 1PN result
(thick, dotted line), to their constrained fit (dashed red line) via Eq.~\eqref{eq:rhofit}, and to the
2PN-accurate (dashed line) and 4PN-accurate (dash-dotted line) approximations of this latter.
In this case, the 2PN, 3PN and 4PN coefficients are given by $c_2^{22}=4.78752$, $c_3^{22}=26.76014$
and $c_4^{22}=43.86148$.
As mentioned above, the important information conveyed by this comparison is that the 1PN
analytical result gives only a rough approximation to the numerical data, since it differs
from them by about a factor of 3 when $x\to 1/3$. Interestingly, the plot shows that one needs
to expand Eq.~\eqref{eq:rhofit} to (at least) 4PN (effective) order to have a good averaged approximation
to the full data. This result also suggests that one would probably need {\it at least}
a 4PN analytical calculation (beyond the leading order $x^9$ contribution) of the absorbed flux
(i.e., up to $(v/c)^{26}$) in the test-mass limit so to have sufficient analytical
information to provide a good approximation to the exact data. Due to recent achievements in
the computation of the $\rho_\lm^\infty(x;,0)$~\cite{Fujita:2011zk}, this seems a doable task.

\section{The horizon fluxes in the $\nu$-dependent case}
\label{sec:horizon_nu}
Let us now turn to discuss the $\nu$-dependent case of $\tilde{\rho}_\lm^H(\x;\,\nu)$
and in particular how to use the $\nu=0$ information there. As mentioned above,
Ref.~\cite{Damour:2008gu} argued that it is meaningful to {\it hybridize}
the $\nu$-dependent with $\nu$-independent information for the $\rho_\lm^\infty$
provided that the $\nu$-dependence of the coefficients of their Taylor
expansion is sufficiently mild.
Following this reasoning, Ref.~\cite{Damour:2008gu} developed the so-called
$3^{+2}$PN approximation to the $\rho_\lm^\infty$: the $\nu$-independent
terms (up to 5PN fractional accuracy for the $\ell=m=2$ multipole) 
added to the $\nu$-dependent ones so to obtain expressions that are formally of
high PN order, though the $\nu$-dependence appears only in the low-order
terms (e.g., up to 3PN for the $\ell=m=2$ multipole).

We aim here at applying the same methodology to increase the usable analytical
information available for the $\tilde{\rho}_\lm^H(\x;\,\nu)$'s. After our analysis
in the test-mass limit, we have clearly seen that the current analytical
approximation for the $\tilde{\rho}_\lm^H$ is rather inaccurate when $\x$ increases.
Luckily, we have also found that, at least for the $\ell=m=2$ case, the $\nu$-dependence
of $\tilde{\rho}_{22}(\x;\,\nu)$ is rather mild, which suggests that an hybridization
procedure can also make sense here. The unfortunate complication is that
the maximum analytical information for the $\tilde{\rho}_\lm^H$ is the same for
both the $\nu=0$ and $\nu\neq 0$ cases. To overcome this problem,
we propose then the following strategy. Since the rational function
representation of $\rho_\lm^H(x;\,0)$ looks robust enough, we use it to
obtain a hopefully reliable estimate of the (effective) PN coefficients up to 4PN.
We then hybridize the 1PN, $\nu$-dependent analytical coefficient with
the so-obtained test-mass 2PN, 3PN and 4PN coefficients.
Focusing on the most relevant $\ell=m=2$ multipole for definiteness,
we propose to write $\tilde{\rho}_{22}^H(\x;\,\nu)$ as the following
Taylor-expanded expression
\begin{align}
\label{eq:rho22H_bb}
\tilde{\rho}_{22}^H(\x;\,\nu) &= 1 + \dfrac{4-18\nu+15\nu^2-2\nu^3}{4(1-4\nu+2\nu^2)}\x \nonumber\\
                    &+ c_2^{22}\x^2 + c_3^{22} \x^3 + c_4^{22}\x^4 ,
\end{align}
where the 1PN coefficient is the exact, $\nu$-dependent one coming from
the analytic calculation of Eq.~\eqref{eq:rho22_anlyt}, while the
($\nu$-independent) $c_i^{22}$ coefficients with $i=2,3,4$ are obtained
from the Taylor expansion of the fitting function to the test-mass numerical
$\rho_{22}^H(x;\,0)$ with the 1PN-constrain $d_1^{22}=1-c_1^{22}$ mentioned in
the previous section.
We finally plug Eq.~\eqref{eq:rho22H_bb} into Eq.~\eqref{eq:FHunstable}
so to complete our proposal to describe the ($\ell=m=2$) horizon flux along the
circular binary dynamics, including unstable orbits.
For simplicity we consider the EOB model with the 3PN-accurate representation of the
$A(u)$ function discussed above, but there is evidently no difficulty in computing the
same quantity with the 5PN-accurate, NR-tuned EOB potential, i.e in the  presence of
the $(a_5(\nu),a_6(\nu))$ functions of Refs.~\cite{Damour:2009kr,Pan:2011gk}.

The two panels of Fig.~\ref{fig:fluxes} compare and contrast the $\ell=m=2$ horizon and asymptotic
fluxes, plotted versus $x=\x$ in the test-mass limit ($\nu=0$, left panel) and versus $\x$
in the equal-mass case ($\nu=1/4$, right-panel).
The blow up, that for the test-mass limit happens at $x^{\rm LR}=1/3$,
occurs at $\x^{\rm LR} =0.3625$  (corresponding to $r_{\rm LR}=2.8456$) in the 3PN-accurate,
equal-mass case. Note that in the test-mass limit we present three different analytical
curves: (i) the resummed asymptotic flux, with
a 5PN-accurate expression for $\rho_{22}^H(x;\,0)$; (ii) the ``exact'' horizon flux, where $\rho_{22}^H(x;\,0)$
is given by a complete fit to a rational function, and (iii) the 4PN approximation
to this latter fit given by Eq.~\eqref{eq:rho22H_bb} above. The figure indicates
that this approximation is rather accurate up to $x\simeq 0.295$, i.e. $r\simeq 3.4$, which gives us confidence that it is meaningful to use it also in the equal-mass case.
In addition we also show for completeness the $F_{22}^{H_{\rm num}}/\nu^2$ numerical points.
In the equal-mass case the horizon flux is compared to the asymptotic flux computed at
the $3^{+2}PN$ approximation from Eq.~\eqref{eq:Finfty_unstable}, notably using the
$\tilde{\rho}_{22}(x;\,\nu)$ of Eq.~\eqref{eq:tilde_rho} (hybridized with 4PN and 5PN
test-mass terms).

The same procedure can be followed for the other multipoles, that will become more and more
relevant for decreasing values of $\nu$. Evidently, since the $\nu$-dependent analytical information is
currently limited to $\tilde{\rho}_{22}(x;\,\nu)$ only, one is obliged to use the test-mass
$\rho_\lm(\x;\,0)$ in PN-expanded form for the subdominant modes. Similarly, concerning
the leading-order prefactor $F_\lm^{(H,\epsilon)}(\x;\,\nu)$, the full $\nu$-dependence is
present only for the $\ell=2$, $m=1$ (see Eq.~\eqref{eq:F21Hpn})
multipole, while for the others one can only use the partial $\nu$-dependent
information provided by Eq.~\eqref{eq:PS94}.

\section{Conclusions}
\label{sec:end}
We have presented a recipe, within the EOB approach, to put together analytical and numerical results so to build an
analytical expression of the horizon-absorbed flux of gravitational waves for circularized
black-hole binaries.
Our result is not limited by the weak-field, slow-motion approximation, but it is 
well-behaved also in the strong-field, fast-velocity regime, notably up to the
last {\it unstable} circular orbit as defined by the EOB adiabatic dynamics.
In this respect, our final analytical expressions incorporate the slow-motion results currently
present in the literature~\cite{Poisson:1994yf,Alvi:2001mx,Poisson:2005pi,Taylor:2008xy},
but go beyond them adding new information in a resummed form.
In practical terms, we do so by introducing a suitable factorization of the horizon flux,
analogous to what was done some time ago for the asymptotic flux~\cite{Damour:2008gu}.
This factorization is such that the only information that needs to be computed is encoded
in the residual amplitude correction to the horizon waveform, $\tilde{\rho}_\lm^H(\x;\,\nu)$, which
plays a role analogous to the residual amplitude corrections to the asymptotic waveform,
$\tilde{\rho}_\lm^\infty(\x;\,\nu)$ introduced in Ref.~\cite{Damour:2008gu} and slightly
modified here to make it well-behaved also below the LSO.

In the test-mass limit ($\nu=0$), we have computed the  $\rho_\lm^H(x;\,0)$'s numerically and we
have then fitted them with certain rational functions so to obtain a portable analytical
representation. The accuracy of such a fit, and in particular the consistency with
1PN-accurate analytical results extracted from Ref.~\cite{Taylor:2008xy} ($\sim 7\%$ or smaller),
makes us confident that it gives a reliable effective representation of the exact analytical
$\rho_\lm^H(x;\,0)$ as it could be obtained from a high-order PN calculation able to bring in
further corrections to the (fractional) 1PN-accurate result of Ref.~\cite{Taylor:2008xy}.
This calculation is not present in the literature, but we recommend it be pushed
{\it at least} up to fractional 4PN order (i.e. 13PN in all) so to hopefully obtain a
reasonable accuracy towards the light-ring for the $\ell=m=2$ mode.

In the $\nu$-dependent case, we have proposed an expression for
$\tilde{\rho}_{22}^H(\x;\,\nu)$ given by Eq.~\eqref{eq:rho22H_bb} that {\it hybridizes}
the 4PN-accurate Taylor expansion of the test-mass fit with the 1PN-accurate calculation
extracted from~\cite{Taylor:2008xy}. This kind of hybridization can be conceptually performed
also for the other multipoles, although  $\nu$-dependent (leading-order) analytical
information is available only for the $\ell=2$, $m=1$ one, Eq.~\eqref{eq:F21Hpn}.
The $\ell=3$ leading-order expressions  are implicitly present in Ref.~\cite{Poisson:2005pi}
and can be made explicit by suitably extending the computation of Ref.~\cite{Taylor:2008xy}.

The main result of this work is to provide analytical formulae for the horizon flux for
any $\nu$ that are {\it resummed} and that incorporate more information than the next-to-leading-order
results of Ref.~\cite{Taylor:2008xy}. The final aim of such expressions is to be incorporated
(as a piece of the radiation-reaction force) in the EOB description of coalescing, nonspinning,
black-hole binaries.  Although for this kind of systems horizon absorption is expected to be small
(since it is a fractional 4PN effect), nonetheless, it will introduce a correction to the phasing
that might be relevant in ongoing comparisons with NR waveforms. As a preliminary exploration of
this effect, we considered the $(a_5,a_6)$-dependent EOB model of Ref.~\cite{Damour:2009ic}
and we added a term of the form ${\cal F}_\varphi^H = -1/(8\pi\Omega)\tilde{F}_{22}^H$
to the standard radiation reaction. As in Ref.~\cite{Baiotti:2011am}, we selected the values
$a_5=-6.37$ and $a_6=50$ that belong to the region in the $(a_5,a_6)$ plane of good agreement
between EOB and numerical-relativity equal-mass waveform found in Ref.~\cite{Damour:2009kr}
(see also Ref.~\cite{Pan:2011gk}), and we focused on an equal-mass binary
$q=m_A/m_B=1$ ($\nu=1/4$) and a $q=3$ binary ($\nu=0.1875$).
For the $q=1$ case, the dephasing due to our additional ${\cal F}_\varphi^H$ is
$\simeq 5\times 10^{-3}$ rad at the light-ring crossing, and thus of the order of
the estimated phase accuracy of state-of-the-art NR simulations,
$\lesssim 10^{-2}$ rad~\cite{Scheel:2008rj,Reisswig:2009us}; on the contrary, for $q=3$
this accumulated dephasing is as large as $\sim 0.1$ rad at light-ring crossing.
To give more weight to this preliminary analysis, before embarking in the study of the
spinning case where absorption effects are expected to be more relevant,
it will be important to  assess the accuracy of the fits to the test-mass
$\rho_{\lm}^H$'s, and thus of the analytical ${\cal F}^H_\varphi$, in the dynamical case of BBH
coalescence in the test-particle limit~\cite{Bernuzzi:2010ty,Bernuzzi:2010xj,Bernuzzi:2011aj}.
This work is currently in progress and will be presented in a separate publication~\cite{Bernuzzi:2012}

\acknowledgments
We thank Leor Barack and Carsten Gundlach for the discussions that inspired this investigation;
AN is grateful to Thibault Damour for important inputs during its development and to Sebastiano
Bernuzzi for a careful reading of the manuscript.

\bibliography{refs20111213}

\begin{thebibliography}{10}%
\makeatletter
\providecommand \@ifxundefined [1]{%
 \ifx #1\undefined \expandafter \@firstoftwo
 \else \expandafter \@secondoftwo
\fi
}%
\providecommand \@ifnum [1]{%
 \ifnum #1\expandafter \@firstoftwo
 \else \expandafter \@secondoftwo
\fi
}%
\providecommand \enquote [1]{``#1''}%
\providecommand \bibnamefont  [1]{#1}%
\providecommand \bibfnamefont [1]{#1}%
\providecommand \citenamefont [1]{#1}%
\providecommand\href[0]{\@sanitize\@href}%
\providecommand\@href[1]{\endgroup\@@startlink{#1}\endgroup\@@href}%
\providecommand\@@href[1]{#1\@@endlink}%
\providecommand \@sanitize [0]{\begingroup\catcode`\&12\catcode`\#12\relax}%
\@ifxundefined \pdfoutput {\@firstoftwo}{%
 \@ifnum{\z@=\pdfoutput}{\@firstoftwo}{\@secondoftwo}%
}{%
 \providecommand\@@startlink[1]{\leavevmode\special{html:<a href="#1">}}%
 \providecommand\@@endlink[0]{\special{html:</a>}}%
}{%
 \providecommand\@@startlink[1]{%
  \leavevmode
  \pdfstartlink
   attr{/Border[0 0 1 ]/H/I/C[0 1 1]}%
   user{/Subtype/Link/A<</Type/Action/S/URI/URI(#1)>>}%
  \relax
 }%
 \providecommand\@@endlink[0]{\pdfendlink}%
}%
\providecommand \url  [0]{\begingroup\@sanitize \@url }%
\providecommand \@url [1]{\endgroup\@href {#1}{\urlprefix}}%
\providecommand \urlprefix [0]{URL }%
\providecommand \Eprint[0]{\href }%
\@ifxundefined \urlstyle {%
  \providecommand \doi [1]{doi:\discretionary{}{}{}#1}%
}{%
  \providecommand \doi [0]{doi:\discretionary{}{}{}\begingroup
  \urlstyle{rm}\Url }%
}%
\providecommand \doibase [0]{http://dx.doi.org/}%
\providecommand \Doi[1]{\href{\doibase#1}}%
\providecommand \bibAnnote [3]{%
  \BibitemShut{#1}%
  \begin{quotation}\noindent
    \textsc{Key:}\ #2\\\textsc{Annotation:}\ #3%
  \end{quotation}%
}%
\providecommand \bibAnnoteFile [2]{%
  \IfFileExists{#2}{\bibAnnote {#1} {#2} {\input{#2}}}{}%
}%
\providecommand \typeout [0]{\immediate \write \m@ne }%
\providecommand \selectlanguage [0]{\@gobble}%
\providecommand \bibinfo [0]{\@secondoftwo}%
\providecommand \bibfield [0]{\@secondoftwo}%
\providecommand \translation [1]{[#1]}%
\providecommand \BibitemOpen[0]{}%
\providecommand \bibitemStop [0]{}%
\providecommand \bibitemNoStop [0]{.\EOS\space}%
\providecommand \EOS [0]{\spacefactor3000\relax}%
\providecommand \BibitemShut [1]{\csname bibitem#1\endcsname}%
\bibitem{Buonanno:1998gg}%
  \BibitemOpen
  \bibfield{author}{%
  \bibinfo {author} {\bibfnamefont{A.}~\bibnamefont{Buonanno}}\ and\ \bibinfo
  {author} {\bibfnamefont{T.}~\bibnamefont{Damour}},\ }%
  \bibfield{journal}{%
  \Doi{10.1103/PhysRevD.59.084006}{\bibinfo {journal} {Phys. Rev.}}\ }%
  \textbf{\bibinfo {volume} {D59}},\ \bibinfo {pages} {084006} (\bibinfo {year}
  {1999})%
  \bibAnnoteFile{NoStop}{Buonanno:1998gg}%
\bibitem{Buonanno:2000ef}%
  \BibitemOpen
  \bibfield{author}{%
  \bibinfo {author} {\bibfnamefont{A.}~\bibnamefont{Buonanno}}\ and\ \bibinfo
  {author} {\bibfnamefont{T.}~\bibnamefont{Damour}},\ }%
  \bibfield{journal}{%
  \Doi{10.1103/PhysRevD.62.064015}{\bibinfo {journal} {Phys. Rev.}}\ }%
  \textbf{\bibinfo {volume} {D62}},\ \bibinfo {pages} {064015} (\bibinfo {year}
  {2000})%
  \bibAnnoteFile{NoStop}{Buonanno:2000ef}%
\bibitem{Damour:2000we}%
  \BibitemOpen
  \bibfield{author}{%
  \bibinfo {author} {\bibfnamefont{T.}~\bibnamefont{Damour}}, \bibinfo {author}
  {\bibfnamefont{P.}~\bibnamefont{Jaranowski}},\ and\ \bibinfo {author}
  {\bibfnamefont{G.}~\bibnamefont{Schaefer}},\ }%
  \bibfield{journal}{%
  \Doi{10.1103/PhysRevD.62.084011}{\bibinfo {journal} {Phys. Rev.}}\ }%
  \textbf{\bibinfo {volume} {D62}},\ \bibinfo {pages} {084011} (\bibinfo {year}
  {2000})%
  \bibAnnoteFile{NoStop}{Damour:2000we}%
\bibitem{Damour:2001tu}%
  \BibitemOpen
  \bibfield{author}{%
  \bibinfo {author} {\bibfnamefont{T.}~\bibnamefont{Damour}},\ }%
  \bibfield{journal}{%
  \Doi{10.1103/PhysRevD.64.124013}{\bibinfo {journal} {Phys. Rev.}}\ }%
  \textbf{\bibinfo {volume} {D64}},\ \bibinfo {pages} {124013} (\bibinfo {year}
  {2001})%
  \bibAnnoteFile{NoStop}{Damour:2001tu}%
\bibitem{Damour:2008gu}%
  \BibitemOpen
  \bibfield{author}{%
  \bibinfo {author} {\bibfnamefont{T.}~\bibnamefont{Damour}}, \bibinfo {author}
  {\bibfnamefont{B.~R.}\ \bibnamefont{Iyer}},\ and\ \bibinfo {author}
  {\bibfnamefont{A.}~\bibnamefont{Nagar}},\ }%
  \bibfield{journal}{%
  \Doi{10.1103/PhysRevD.79.064004}{\bibinfo {journal} {Phys. Rev.}}\ }%
  \textbf{\bibinfo {volume} {D79}},\ \bibinfo {pages} {064004} (\bibinfo {year}
  {2009})%
  \bibAnnoteFile{NoStop}{Damour:2008gu}%
\bibitem{Damour:2008qf}%
  \BibitemOpen
  \bibfield{author}{%
  \bibinfo {author} {\bibfnamefont{T.}~\bibnamefont{Damour}}, \bibinfo {author}
  {\bibfnamefont{P.}~\bibnamefont{Jaranowski}},\ and\ \bibinfo {author}
  {\bibfnamefont{G.}~\bibnamefont{Schaefer}},\ }%
  \bibfield{journal}{%
  \Doi{10.1103/PhysRevD.78.024009}{\bibinfo {journal} {Phys.Rev.}}\ }%
  \textbf{\bibinfo {volume} {D78}},\ \bibinfo {pages} {024009} (\bibinfo {year}
  {2008}),\ \Eprint{http://arxiv.org/abs/0803.0915}{arXiv:0803.0915 [gr-qc]}%
  \bibAnnoteFile{NoStop}{Damour:2008qf}%
\bibitem{Damour:2009wj}%
  \BibitemOpen
  \bibfield{author}{%
  \bibinfo {author} {\bibfnamefont{T.}~\bibnamefont{Damour}}\ and\ \bibinfo
  {author} {\bibfnamefont{A.}~\bibnamefont{Nagar}},\ }%
  \bibfield{journal}{%
  \Doi{10.1103/PhysRevD.81.084016}{\bibinfo {journal} {Phys.Rev.}}\ }%
  \textbf{\bibinfo {volume} {D81}},\ \bibinfo {pages} {084016} (\bibinfo {year}
  {2010}),\ \Eprint{http://arxiv.org/abs/0911.5041}{arXiv:0911.5041 [gr-qc]}%
  \bibAnnoteFile{NoStop}{Damour:2009wj}%
\bibitem{Damour:2009ic}%
  \BibitemOpen
  \bibfield{author}{%
  \bibinfo {author} {\bibfnamefont{T.}~\bibnamefont{Damour}}\ and\ \bibinfo
  {author} {\bibfnamefont{A.}~\bibnamefont{Nagar}}}%
   (\bibinfo {year} {2009}),\
  \Eprint{http://arxiv.org/abs/0906.1769}{arXiv:0906.1769 [gr-qc]}%
  \bibAnnoteFile{NoStop}{Damour:2009ic}%
\bibitem{Barausse:2009xi}%
  \BibitemOpen
  \bibfield{author}{%
  \bibinfo {author} {\bibfnamefont{E.}~\bibnamefont{Barausse}}\ and\ \bibinfo
  {author} {\bibfnamefont{A.}~\bibnamefont{Buonanno}},\ }%
  \bibfield{journal}{%
  \Doi{10.1103/PhysRevD.81.084024}{\bibinfo {journal} {Phys.Rev.}}\ }%
  \textbf{\bibinfo {volume} {D81}},\ \bibinfo {pages} {084024} (\bibinfo {year}
  {2010}),\ \Eprint{http://arxiv.org/abs/0912.3517}{arXiv:0912.3517 [gr-qc]}%
  \bibAnnoteFile{NoStop}{Barausse:2009xi}%
\bibitem{Pan:2010hz}%
  \BibitemOpen
  \bibfield{author}{%
  \bibinfo {author} {\bibfnamefont{Y.}~\bibnamefont{Pan}}, \bibinfo {author}
  {\bibfnamefont{A.}~\bibnamefont{Buonanno}}, \bibinfo {author}
  {\bibfnamefont{R.}~\bibnamefont{Fujita}}, \bibinfo {author}
  {\bibfnamefont{E.}~\bibnamefont{Racine}},\ and\ \bibinfo {author}
  {\bibfnamefont{H.}~\bibnamefont{Tagoshi}},\ }%
  \bibfield{journal}{%
  \Doi{10.1103/PhysRevD.83.064003}{\bibinfo {journal} {Phys.Rev.}}\ }%
  \textbf{\bibinfo {volume} {D83}},\ \bibinfo {pages} {064003} (\bibinfo {year}
  {2011}),\ \Eprint{http://arxiv.org/abs/1006.0431}{arXiv:1006.0431 [gr-qc]}%
  \bibAnnoteFile{NoStop}{Pan:2010hz}%
\bibitem{Barausse:2011ys}%
  \BibitemOpen
  \bibfield{author}{%
  \bibinfo {author} {\bibfnamefont{E.}~\bibnamefont{Barausse}}\ and\ \bibinfo
  {author} {\bibfnamefont{A.}~\bibnamefont{Buonanno}}}%
   (\bibinfo {year} {2011}),\ \bibinfo {note} {* Temporary entry *},\
  \Eprint{http://arxiv.org/abs/1107.2904}{arXiv:1107.2904 [gr-qc]}%
  \bibAnnoteFile{NoStop}{Barausse:2011ys}%
\bibitem{Barausse:2011dq}%
  \BibitemOpen
  \bibfield{author}{%
  \bibinfo {author} {\bibfnamefont{E.}~\bibnamefont{Barausse}}, \bibinfo
  {author} {\bibfnamefont{A.}~\bibnamefont{Buonanno}},\ and\ \bibinfo {author}
  {\bibfnamefont{A.}~\bibnamefont{Le~Tiec}}}%
   (\bibinfo {year} {2011}),\ \bibinfo {note} {* Temporary entry *},\
  \Eprint{http://arxiv.org/abs/1111.5610}{arXiv:1111.5610 [gr-qc]}%
  \bibAnnoteFile{NoStop}{Barausse:2011dq}%
\bibitem{Pan:2009wj}%
  \BibitemOpen
  \bibfield{author}{%
  \bibinfo {author} {\bibfnamefont{Y.}~\bibnamefont{Pan}}, \bibinfo {author}
  {\bibfnamefont{A.}~\bibnamefont{Buonanno}}, \bibinfo {author}
  {\bibfnamefont{L.~T.}\ \bibnamefont{Buchman}}, \bibinfo {author}
  {\bibfnamefont{T.}~\bibnamefont{Chu}}, \bibinfo {author}
  {\bibfnamefont{L.~E.}\ \bibnamefont{Kidder}}, \emph{et~al.},\ }%
  \bibfield{journal}{%
  \Doi{10.1103/PhysRevD.81.084041}{\bibinfo {journal} {Phys.Rev.}}\ }%
  \textbf{\bibinfo {volume} {D81}},\ \bibinfo {pages} {084041} (\bibinfo {year}
  {2010}),\ \Eprint{http://arxiv.org/abs/0912.3466}{arXiv:0912.3466 [gr-qc]}%
  \bibAnnoteFile{NoStop}{Pan:2009wj}%
\bibitem{Baiotti:2010xh}%
  \BibitemOpen
  \bibfield{author}{%
  \bibinfo {author} {\bibfnamefont{L.}~\bibnamefont{Baiotti}}, \bibinfo
  {author} {\bibfnamefont{T.}~\bibnamefont{Damour}}, \bibinfo {author}
  {\bibfnamefont{B.}~\bibnamefont{Giacomazzo}}, \bibinfo {author}
  {\bibfnamefont{A.}~\bibnamefont{Nagar}},\ and\ \bibinfo {author}
  {\bibfnamefont{L.}~\bibnamefont{Rezzolla}},\ }%
  \bibfield{journal}{%
  \Doi{10.1103/PhysRevLett.105.261101}{\bibinfo {journal} {Phys.Rev.Lett.}}\ }%
  \textbf{\bibinfo {volume} {105}},\ \bibinfo {pages} {261101} (\bibinfo {year}
  {2010}),\ \Eprint{http://arxiv.org/abs/1009.0521}{arXiv:1009.0521 [gr-qc]}%
  \bibAnnoteFile{NoStop}{Baiotti:2010xh}%
\bibitem{Baiotti:2011am}%
  \BibitemOpen
  \bibfield{author}{%
  \bibinfo {author} {\bibfnamefont{L.}~\bibnamefont{Baiotti}}, \bibinfo
  {author} {\bibfnamefont{T.}~\bibnamefont{Damour}}, \bibinfo {author}
  {\bibfnamefont{B.}~\bibnamefont{Giacomazzo}}, \bibinfo {author}
  {\bibfnamefont{A.}~\bibnamefont{Nagar}},\ and\ \bibinfo {author}
  {\bibfnamefont{L.}~\bibnamefont{Rezzolla}},\ }%
  \bibfield{journal}{%
  \Doi{10.1103/PhysRevD.84.024017}{\bibinfo {journal} {Phys.Rev.}}\ }%
  \textbf{\bibinfo {volume} {D84}},\ \bibinfo {pages} {024017} (\bibinfo {year}
  {2011}),\ \Eprint{http://arxiv.org/abs/1103.3874}{arXiv:1103.3874 [gr-qc]}%
  \bibAnnoteFile{NoStop}{Baiotti:2011am}%
\bibitem{Pan:2011gk}%
  \BibitemOpen
  \bibfield{author}{%
  \bibinfo {author} {\bibfnamefont{Y.}~\bibnamefont{Pan}} \emph{et~al.}}%
   (\bibinfo {year} {2011}),\
  \Eprint{http://arxiv.org/abs/1106.1021}{arXiv:1106.1021 [gr-qc]}%
  \bibAnnoteFile{NoStop}{Pan:2011gk}%
\bibitem{Damour:2011fu}%
  \BibitemOpen
  \bibfield{author}{%
  \bibinfo {author} {\bibfnamefont{T.}~\bibnamefont{Damour}}, \bibinfo {author}
  {\bibfnamefont{A.}~\bibnamefont{Nagar}}, \bibinfo {author}
  {\bibfnamefont{D.}~\bibnamefont{Pollney}},\ and\ \bibinfo {author}
  {\bibfnamefont{C.}~\bibnamefont{Reisswig}}}%
   (\bibinfo {year} {2011}),\ \bibinfo {note} {* Temporary entry *},\
  \Eprint{http://arxiv.org/abs/1110.2938}{arXiv:1110.2938 [gr-qc]}%
  \bibAnnoteFile{NoStop}{Damour:2011fu}%
\bibitem{Barack:2010ny}%
  \BibitemOpen
  \bibfield{author}{%
  \bibinfo {author} {\bibfnamefont{L.}~\bibnamefont{Barack}}, \bibinfo {author}
  {\bibfnamefont{T.}~\bibnamefont{Damour}},\ and\ \bibinfo {author}
  {\bibfnamefont{N.}~\bibnamefont{Sago}},\ }%
  \bibfield{journal}{%
  \Doi{10.1103/PhysRevD.82.084036}{\bibinfo {journal} {Phys.Rev.}}\ }%
  \textbf{\bibinfo {volume} {D82}},\ \bibinfo {pages} {084036} (\bibinfo {year}
  {2010}),\ \Eprint{http://arxiv.org/abs/1008.0935}{arXiv:1008.0935 [gr-qc]}%
  \bibAnnoteFile{NoStop}{Barack:2010ny}%
\bibitem{Damour:2010zb}%
  \BibitemOpen
  \bibfield{author}{%
  \bibinfo {author} {\bibfnamefont{T.}~\bibnamefont{Damour}}, \bibinfo {author}
  {\bibfnamefont{A.}~\bibnamefont{Nagar}},\ and\ \bibinfo {author}
  {\bibfnamefont{M.}~\bibnamefont{Trias}},\ }%
  \bibfield{journal}{%
  \Doi{10.1103/PhysRevD.83.024006}{\bibinfo {journal} {Phys.Rev.}}\ }%
  \textbf{\bibinfo {volume} {D83}},\ \bibinfo {pages} {024006} (\bibinfo {year}
  {2011}),\ \Eprint{http://arxiv.org/abs/1009.5998}{arXiv:1009.5998 [gr-qc]}%
  \bibAnnoteFile{NoStop}{Damour:2010zb}%
\bibitem{Poisson:1994yf}%
  \BibitemOpen
  \bibfield{author}{%
  \bibinfo {author} {\bibfnamefont{E.}~\bibnamefont{Poisson}}\ and\ \bibinfo
  {author} {\bibfnamefont{M.}~\bibnamefont{Sasaki}},\ }%
  \bibfield{journal}{%
  \Doi{10.1103/PhysRevD.51.5753}{\bibinfo {journal} {Phys.Rev.}}\ }%
  \textbf{\bibinfo {volume} {D51}},\ \bibinfo {pages} {5753} (\bibinfo {year}
  {1995}),\ \Eprint{http://arxiv.org/abs/gr-qc/9412027}{arXiv:gr-qc/9412027
  [gr-qc]}%
  \bibAnnoteFile{NoStop}{Poisson:1994yf}%
\bibitem{Tagoshi:1997jy}%
  \BibitemOpen
  \bibfield{author}{%
  \bibinfo {author} {\bibfnamefont{H.}~\bibnamefont{Tagoshi}}, \bibinfo
  {author} {\bibfnamefont{S.}~\bibnamefont{Mano}},\ and\ \bibinfo {author}
  {\bibfnamefont{E.}~\bibnamefont{Takasugi}},\ }%
  \bibfield{journal}{%
  \Doi{10.1143/PTP.98.829}{\bibinfo {journal} {Prog.Theor.Phys.}}\ }%
  \textbf{\bibinfo {volume} {98}},\ \bibinfo {pages} {829} (\bibinfo {year}
  {1997}),\ \Eprint{http://arxiv.org/abs/gr-qc/9711072}{arXiv:gr-qc/9711072
  [gr-qc]}%
  \bibAnnoteFile{NoStop}{Tagoshi:1997jy}%
\bibitem{Alvi:2001mx}%
  \BibitemOpen
  \bibfield{author}{%
  \bibinfo {author} {\bibfnamefont{K.}~\bibnamefont{Alvi}},\ }%
  \bibfield{journal}{%
  \Doi{10.1103/PhysRevD.64.104020}{\bibinfo {journal} {Phys.Rev.}}\ }%
  \textbf{\bibinfo {volume} {D64}},\ \bibinfo {pages} {104020} (\bibinfo {year}
  {2001}),\ \Eprint{http://arxiv.org/abs/gr-qc/0107080}{arXiv:gr-qc/0107080
  [gr-qc]}%
  \bibAnnoteFile{NoStop}{Alvi:2001mx}%
\bibitem{Poisson:2005pi}%
  \BibitemOpen
  \bibfield{author}{%
  \bibinfo {author} {\bibfnamefont{E.}~\bibnamefont{Poisson}},\ }%
  \bibfield{journal}{%
  \Doi{10.1103/PhysRevLett.94.161103}{\bibinfo {journal} {Phys.Rev.Lett.}}\ }%
  \textbf{\bibinfo {volume} {94}},\ \bibinfo {pages} {161103} (\bibinfo {year}
  {2005}),\ \Eprint{http://arxiv.org/abs/gr-qc/0501032}{arXiv:gr-qc/0501032
  [gr-qc]}%
  \bibAnnoteFile{NoStop}{Poisson:2005pi}%
\bibitem{Taylor:2008xy}%
  \BibitemOpen
  \bibfield{author}{%
  \bibinfo {author} {\bibfnamefont{S.}~\bibnamefont{Taylor}}\ and\ \bibinfo
  {author} {\bibfnamefont{E.}~\bibnamefont{Poisson}},\ }%
  \bibfield{journal}{%
  \Doi{10.1103/PhysRevD.78.084016}{\bibinfo {journal} {Phys.Rev.}}\ }%
  \textbf{\bibinfo {volume} {D78}},\ \bibinfo {pages} {084016} (\bibinfo {year}
  {2008}),\ \Eprint{http://arxiv.org/abs/0806.3052}{arXiv:0806.3052 [gr-qc]}%
  \bibAnnoteFile{NoStop}{Taylor:2008xy}%
\bibitem{Comeau:2009bz}%
  \BibitemOpen
  \bibfield{author}{%
  \bibinfo {author} {\bibfnamefont{S.}~\bibnamefont{Comeau}}\ and\ \bibinfo
  {author} {\bibfnamefont{E.}~\bibnamefont{Poisson}},\ }%
  \bibfield{journal}{%
  \Doi{10.1103/PhysRevD.80.087501}{\bibinfo {journal} {Phys.Rev.}}\ }%
  \textbf{\bibinfo {volume} {D80}},\ \bibinfo {pages} {087501} (\bibinfo {year}
  {2009}),\ \Eprint{http://arxiv.org/abs/0908.4518}{arXiv:0908.4518 [gr-qc]}%
  \bibAnnoteFile{NoStop}{Comeau:2009bz}%
\bibitem{Poisson:2009qj}%
  \BibitemOpen
  \bibfield{author}{%
  \bibinfo {author} {\bibfnamefont{E.}~\bibnamefont{Poisson}}\ and\ \bibinfo
  {author} {\bibfnamefont{I.}~\bibnamefont{Vlasov}},\ }%
  \bibfield{journal}{%
  \Doi{10.1103/PhysRevD.81.024029}{\bibinfo {journal} {Phys.Rev.}}\ }%
  \textbf{\bibinfo {volume} {D81}},\ \bibinfo {pages} {024029} (\bibinfo {year}
  {2010}),\ \Eprint{http://arxiv.org/abs/0910.4311}{arXiv:0910.4311 [gr-qc]}%
  \bibAnnoteFile{NoStop}{Poisson:2009qj}%
\bibitem{Price:2001un}%
  \BibitemOpen
  \bibfield{author}{%
  \bibinfo {author} {\bibfnamefont{R.~H.}\ \bibnamefont{Price}}\ and\ \bibinfo
  {author} {\bibfnamefont{J.~T.}\ \bibnamefont{Whelan}},\ }%
  \bibfield{journal}{%
  \Doi{10.1103/PhysRevLett.87.231101}{\bibinfo {journal} {Phys.Rev.Lett.}}\ }%
  \textbf{\bibinfo {volume} {87}},\ \bibinfo {pages} {231101} (\bibinfo {year}
  {2001}),\ \Eprint{http://arxiv.org/abs/gr-qc/0107029}{arXiv:gr-qc/0107029
  [gr-qc]}%
  \bibAnnoteFile{NoStop}{Price:2001un}%
\bibitem{Scheel:2008rj}%
  \BibitemOpen
  \bibfield{author}{%
  \bibinfo {author} {\bibfnamefont{M.~A.}\ \bibnamefont{Scheel}}
  \emph{et~al.},\ }%
  \bibfield{journal}{%
  \Doi{10.1103/PhysRevD.79.024003}{\bibinfo {journal} {Phys. Rev.}}\ }%
  \textbf{\bibinfo {volume} {D79}},\ \bibinfo {pages} {024003} (\bibinfo {year}
  {2009}),\ \Eprint{http://arxiv.org/abs/0810.1767}{arXiv:0810.1767 [gr-qc]}%
  \bibAnnoteFile{NoStop}{Scheel:2008rj}%
\bibitem{Reisswig:2009us}%
  \BibitemOpen
  \bibfield{author}{%
  \bibinfo {author} {\bibfnamefont{C.}~\bibnamefont{Reisswig}} \emph{et~al.},\
  }%
  \bibfield{journal}{%
  \Doi{10.1103/PhysRevLett.103.221101}{\bibinfo {journal} {Phys.Rev.Lett.}}\ }%
  \textbf{\bibinfo {volume} {103}},\ \bibinfo {pages} {221101} (\bibinfo {year}
  {2009})%
  \bibAnnoteFile{NoStop}{Reisswig:2009us}%
\bibitem{Lovelace:2011nu}%
  \BibitemOpen
  \bibfield{author}{%
  \bibinfo {author} {\bibfnamefont{G.}~\bibnamefont{Lovelace}}, \bibinfo
  {author} {\bibfnamefont{M.}~\bibnamefont{Boyle}}, \bibinfo {author}
  {\bibfnamefont{M.~A.}\ \bibnamefont{Scheel}},\ and\ \bibinfo {author}
  {\bibfnamefont{B.}~\bibnamefont{Szilagyi}}}%
   (\bibinfo {year} {2011}),\
  \Eprint{http://arxiv.org/abs/1110.2229}{arXiv:1110.2229 [gr-qc]}%
  \bibAnnoteFile{NoStop}{Lovelace:2011nu}%
\bibitem{Yunes:2010zj}%
  \BibitemOpen
  \bibfield{author}{%
  \bibinfo {author} {\bibfnamefont{N.}~\bibnamefont{Yunes}}, \bibinfo {author}
  {\bibfnamefont{A.}~\bibnamefont{Buonanno}}, \bibinfo {author}
  {\bibfnamefont{S.~A.}\ \bibnamefont{Hughes}}, \bibinfo {author}
  {\bibfnamefont{Y.}~\bibnamefont{Pan}}, \bibinfo {author}
  {\bibfnamefont{E.}~\bibnamefont{Barausse}}, \emph{et~al.},\ }%
  \bibfield{journal}{%
  \Doi{10.1103/PhysRevD.83.044044}{\bibinfo {journal} {Phys.Rev.}}\ }%
  \textbf{\bibinfo {volume} {D83}},\ \bibinfo {pages} {044044} (\bibinfo {year}
  {2011}),\ \Eprint{http://arxiv.org/abs/1009.6013}{arXiv:1009.6013 [gr-qc]}%
  \bibAnnoteFile{NoStop}{Yunes:2010zj}%
\bibitem{Yunes:2009ef}%
  \BibitemOpen
  \bibfield{author}{%
  \bibinfo {author} {\bibfnamefont{N.}~\bibnamefont{Yunes}}, \bibinfo {author}
  {\bibfnamefont{A.}~\bibnamefont{Buonanno}}, \bibinfo {author}
  {\bibfnamefont{S.~A.}\ \bibnamefont{Hughes}}, \bibinfo {author}
  {\bibfnamefont{M.}~\bibnamefont{Coleman~Miller}},\ and\ \bibinfo {author}
  {\bibfnamefont{Y.}~\bibnamefont{Pan}},\ }%
  \bibfield{journal}{%
  \Doi{10.1103/PhysRevLett.104.091102}{\bibinfo {journal} {Phys. Rev. Lett.}}\
  }%
  \textbf{\bibinfo {volume} {104}},\ \bibinfo {pages} {091102} (\bibinfo {year}
  {2010}),\ \Eprint{http://arxiv.org/abs/0909.4263}{arXiv:0909.4263 [gr-qc]}%
  \bibAnnoteFile{NoStop}{Yunes:2009ef}%
\bibitem{Hughes:2001jr}%
  \BibitemOpen
  \bibfield{author}{%
  \bibinfo {author} {\bibfnamefont{S.~A.}\ \bibnamefont{Hughes}},\ }%
  \bibfield{journal}{%
  \Doi{10.1103/PhysRevD.64.064004}{\bibinfo {journal} {Phys.Rev.}}\ }%
  \textbf{\bibinfo {volume} {D64}},\ \bibinfo {pages} {064004} (\bibinfo {year}
  {2001}),\ \Eprint{http://arxiv.org/abs/gr-qc/0104041}{arXiv:gr-qc/0104041
  [gr-qc]}%
  \bibAnnoteFile{NoStop}{Hughes:2001jr}%
\bibitem{Martel:2003jj}%
  \BibitemOpen
  \bibfield{author}{%
  \bibinfo {author} {\bibfnamefont{K.}~\bibnamefont{Martel}},\ }%
  \bibfield{journal}{%
  \Doi{10.1103/PhysRevD.69.044025}{\bibinfo {journal} {Phys.Rev.}}\ }%
  \textbf{\bibinfo {volume} {D69}},\ \bibinfo {pages} {044025} (\bibinfo {year}
  {2004}),\ \Eprint{http://arxiv.org/abs/gr-qc/0311017}{arXiv:gr-qc/0311017
  [gr-qc]}%
  \bibAnnoteFile{NoStop}{Martel:2003jj}%
\bibitem{Barack:2010tm}%
  \BibitemOpen
  \bibfield{author}{%
  \bibinfo {author} {\bibfnamefont{L.}~\bibnamefont{Barack}}\ and\ \bibinfo
  {author} {\bibfnamefont{N.}~\bibnamefont{Sago}},\ }%
  \bibfield{journal}{%
  \Doi{10.1103/PhysRevD.81.084021}{\bibinfo {journal} {Phys. Rev.}}\ }%
  \textbf{\bibinfo {volume} {D81}},\ \bibinfo {pages} {084021} (\bibinfo {year}
  {2010}),\ \Eprint{http://arxiv.org/abs/1002.2386}{arXiv:1002.2386 [gr-qc]}%
  \bibAnnoteFile{NoStop}{Barack:2010tm}%
\bibitem{Vega:2011ue}%
  \BibitemOpen
  \bibfield{author}{%
  \bibinfo {author} {\bibfnamefont{I.}~\bibnamefont{Vega}}, \bibinfo {author}
  {\bibfnamefont{E.}~\bibnamefont{Poisson}},\ and\ \bibinfo {author}
  {\bibfnamefont{R.}~\bibnamefont{Massey}},\ }%
  \bibfield{journal}{%
  \Doi{10.1088/0264-9381/28/17/175006}{\bibinfo {journal} {Class.Quant.Grav.}}\
  }%
  \textbf{\bibinfo {volume} {28}},\ \bibinfo {pages} {175006} (\bibinfo {year}
  {2011}),\ \bibinfo {note} {* Temporary entry *},\
  \Eprint{http://arxiv.org/abs/1106.0510}{arXiv:1106.0510 [gr-qc]}%
  \bibAnnoteFile{NoStop}{Vega:2011ue}%
\bibitem{Pretorius:2007jn}%
  \BibitemOpen
  \bibfield{author}{%
  \bibinfo {author} {\bibfnamefont{F.}~\bibnamefont{Pretorius}}\ and\ \bibinfo
  {author} {\bibfnamefont{D.}~\bibnamefont{Khurana}},\ }%
  \bibfield{journal}{%
  \Doi{10.1088/0264-9381/24/12/S07}{\bibinfo {journal} {Class.Quant.Grav.}}\ }%
  \textbf{\bibinfo {volume} {24}},\ \bibinfo {pages} {S83} (\bibinfo {year}
  {2007}),\ \Eprint{http://arxiv.org/abs/gr-qc/0702084}{arXiv:gr-qc/0702084
  [GR-QC]}%
  \bibAnnoteFile{NoStop}{Pretorius:2007jn}%
\bibitem{Healy:2009zm}%
  \BibitemOpen
  \bibfield{author}{%
  \bibinfo {author} {\bibfnamefont{J.}~\bibnamefont{Healy}}, \bibinfo {author}
  {\bibfnamefont{J.}~\bibnamefont{Levin}},\ and\ \bibinfo {author}
  {\bibfnamefont{D.}~\bibnamefont{Shoemaker}},\ }%
  \bibfield{journal}{%
  \Doi{10.1103/PhysRevLett.103.131101}{\bibinfo {journal} {Phys.Rev.Lett.}}\ }%
  \textbf{\bibinfo {volume} {103}},\ \bibinfo {pages} {131101} (\bibinfo {year}
  {2009}),\ \Eprint{http://arxiv.org/abs/0907.0671}{arXiv:0907.0671 [gr-qc]}%
  \bibAnnoteFile{NoStop}{Healy:2009zm}%
\bibitem{Sperhake:2009jz}%
  \BibitemOpen
  \bibfield{author}{%
  \bibinfo {author} {\bibfnamefont{U.}~\bibnamefont{Sperhake}}, \bibinfo
  {author} {\bibfnamefont{V.}~\bibnamefont{Cardoso}}, \bibinfo {author}
  {\bibfnamefont{F.}~\bibnamefont{Pretorius}}, \bibinfo {author}
  {\bibfnamefont{E.}~\bibnamefont{Berti}}, \bibinfo {author}
  {\bibfnamefont{T.}~\bibnamefont{Hinderer}}, \emph{et~al.},\ }%
  \bibfield{journal}{%
  \Doi{10.1103/PhysRevLett.103.131102}{\bibinfo {journal} {Phys.Rev.Lett.}}\ }%
  \textbf{\bibinfo {volume} {103}},\ \bibinfo {pages} {131102} (\bibinfo {year}
  {2009}),\ \Eprint{http://arxiv.org/abs/0907.1252}{arXiv:0907.1252 [gr-qc]}%
  \bibAnnoteFile{NoStop}{Sperhake:2009jz}%
\bibitem{Sperhake:2010uv}%
  \BibitemOpen
  \bibfield{author}{%
  \bibinfo {author} {\bibfnamefont{U.}~\bibnamefont{Sperhake}}, \bibinfo
  {author} {\bibfnamefont{E.}~\bibnamefont{Berti}}, \bibinfo {author}
  {\bibfnamefont{V.}~\bibnamefont{Cardoso}}, \bibinfo {author}
  {\bibfnamefont{F.}~\bibnamefont{Pretorius}},\ and\ \bibinfo {author}
  {\bibfnamefont{N.}~\bibnamefont{Yunes}},\ }%
  \bibfield{journal}{%
  \Doi{10.1103/PhysRevD.83.024037}{\bibinfo {journal} {Phys.Rev.}}\ }%
  \textbf{\bibinfo {volume} {D83}},\ \bibinfo {pages} {024037} (\bibinfo {year}
  {2011}),\ \Eprint{http://arxiv.org/abs/1011.3281}{arXiv:1011.3281 [gr-qc]}%
  \bibAnnoteFile{NoStop}{Sperhake:2010uv}%
\bibitem{Damour:2009kr}%
  \BibitemOpen
  \bibfield{author}{%
  \bibinfo {author} {\bibfnamefont{T.}~\bibnamefont{Damour}}\ and\ \bibinfo
  {author} {\bibfnamefont{A.}~\bibnamefont{Nagar}},\ }%
  \bibfield{journal}{%
  \Doi{10.1103/PhysRevD.79.081503}{\bibinfo {journal} {Phys. Rev.}}\ }%
  \textbf{\bibinfo {volume} {D79}},\ \bibinfo {pages} {081503} (\bibinfo {year}
  {2009})%
  \bibAnnoteFile{NoStop}{Damour:2009kr}%
\bibitem{Poisson:2004cw}%
  \BibitemOpen
  \bibfield{author}{%
  \bibinfo {author} {\bibfnamefont{E.}~\bibnamefont{Poisson}},\ }%
  \bibfield{journal}{%
  \Doi{10.1103/PhysRevD.70.084044}{\bibinfo {journal} {Phys.Rev.}}\ }%
  \textbf{\bibinfo {volume} {D70}},\ \bibinfo {pages} {084044} (\bibinfo {year}
  {2004}),\ \Eprint{http://arxiv.org/abs/gr-qc/0407050}{arXiv:gr-qc/0407050
  [gr-qc]}%
  \bibAnnoteFile{NoStop}{Poisson:2004cw}%
\bibitem{Fujita:2011zk}%
  \BibitemOpen
  \bibfield{author}{%
  \bibinfo {author} {\bibfnamefont{R.}~\bibnamefont{Fujita}}}%
   (\bibinfo {year} {2011}),\
  \Eprint{http://arxiv.org/abs/1104.5615}{arXiv:1104.5615 [gr-qc]}%
  \bibAnnoteFile{NoStop}{Fujita:2011zk}%
\bibitem{Mano:1996vt}%
  \BibitemOpen
  \bibfield{author}{%
  \bibinfo {author} {\bibfnamefont{S.}~\bibnamefont{Mano}}, \bibinfo {author}
  {\bibfnamefont{H.}~\bibnamefont{Suzuki}},\ and\ \bibinfo {author}
  {\bibfnamefont{E.}~\bibnamefont{Takasugi}},\ }%
  \bibfield{journal}{%
  \Doi{10.1143/PTP.95.1079}{\bibinfo {journal} {Prog.Theor.Phys.}}\ }%
  \textbf{\bibinfo {volume} {95}},\ \bibinfo {pages} {1079} (\bibinfo {year}
  {1996}),\ \Eprint{http://arxiv.org/abs/gr-qc/9603020}{arXiv:gr-qc/9603020
  [gr-qc]}%
  \bibAnnoteFile{NoStop}{Mano:1996vt}%
\bibitem{Tanaka:1997dj}%
  \BibitemOpen
  \bibfield{author}{%
  \bibinfo {author} {\bibfnamefont{T.}~\bibnamefont{Tanaka}}, \bibinfo {author}
  {\bibfnamefont{H.}~\bibnamefont{Tagoshi}},\ and\ \bibinfo {author}
  {\bibfnamefont{M.}~\bibnamefont{Sasaki}},\ }%
  \bibfield{journal}{%
  \Doi{10.1143/PTP.96.1087}{\bibinfo {journal} {Prog.Theor.Phys.}}\ }%
  \textbf{\bibinfo {volume} {96}},\ \bibinfo {pages} {1087} (\bibinfo {year}
  {1996}),\ \bibinfo {note} {revised},\
  \Eprint{http://arxiv.org/abs/gr-qc/9701050}{arXiv:gr-qc/9701050 [gr-qc]}%
  \bibAnnoteFile{NoStop}{Tanaka:1997dj}%
\bibitem{Blanchet:2001aw}%
  \BibitemOpen
  \bibfield{author}{%
  \bibinfo {author} {\bibfnamefont{L.}~\bibnamefont{Blanchet}}, \bibinfo
  {author} {\bibfnamefont{B.~R.}\ \bibnamefont{Iyer}},\ and\ \bibinfo {author}
  {\bibfnamefont{B.}~\bibnamefont{Joguet}},\ }%
  \bibfield{journal}{%
  \Doi{10.1103/PhysRevD.65.064005, 10.1103/PhysRevD.71.129903}{\bibinfo
  {journal} {Phys.Rev.}}\ }%
  \textbf{\bibinfo {volume} {D65}},\ \bibinfo {pages} {064005} (\bibinfo {year}
  {2002}),\ \Eprint{http://arxiv.org/abs/gr-qc/0105098}{arXiv:gr-qc/0105098
  [gr-qc]}%
  \bibAnnoteFile{NoStop}{Blanchet:2001aw}%
\bibitem{Blanchet:2001ax}%
  \BibitemOpen
  \bibfield{author}{%
  \bibinfo {author} {\bibfnamefont{L.}~\bibnamefont{Blanchet}}, \bibinfo
  {author} {\bibfnamefont{G.}~\bibnamefont{Faye}}, \bibinfo {author}
  {\bibfnamefont{B.~R.}\ \bibnamefont{Iyer}},\ and\ \bibinfo {author}
  {\bibfnamefont{B.}~\bibnamefont{Joguet}},\ }%
  \bibfield{journal}{%
  \Doi{10.1103/PhysRevD.65.061501, 10.1103/PhysRevD.71.129902}{\bibinfo
  {journal} {Phys.Rev.}}\ }%
  \textbf{\bibinfo {volume} {D65}},\ \bibinfo {pages} {061501} (\bibinfo {year}
  {2002}),\ \Eprint{http://arxiv.org/abs/gr-qc/0105099}{arXiv:gr-qc/0105099
  [gr-qc]}%
  \bibAnnoteFile{NoStop}{Blanchet:2001ax}%
\bibitem{Blanchet:2004ek}%
  \BibitemOpen
  \bibfield{author}{%
  \bibinfo {author} {\bibfnamefont{L.}~\bibnamefont{Blanchet}}, \bibinfo
  {author} {\bibfnamefont{T.}~\bibnamefont{Damour}}, \bibinfo {author}
  {\bibfnamefont{G.}~\bibnamefont{Esposito-Farese}},\ and\ \bibinfo {author}
  {\bibfnamefont{B.~R.}\ \bibnamefont{Iyer}},\ }%
  \bibfield{journal}{%
  \Doi{10.1103/PhysRevLett.93.091101}{\bibinfo {journal} {Phys.Rev.Lett.}}\ }%
  \textbf{\bibinfo {volume} {93}},\ \bibinfo {pages} {091101} (\bibinfo {year}
  {2004}),\ \Eprint{http://arxiv.org/abs/gr-qc/0406012}{arXiv:gr-qc/0406012
  [gr-qc]}%
  \bibAnnoteFile{NoStop}{Blanchet:2004ek}%
\bibitem{Blanchet:2005tk}%
  \BibitemOpen
  \bibfield{author}{%
  \bibinfo {author} {\bibfnamefont{L.}~\bibnamefont{Blanchet}}, \bibinfo
  {author} {\bibfnamefont{T.}~\bibnamefont{Damour}}, \bibinfo {author}
  {\bibfnamefont{G.}~\bibnamefont{Esposito-Farese}},\ and\ \bibinfo {author}
  {\bibfnamefont{B.~R.}\ \bibnamefont{Iyer}},\ }%
  \bibfield{journal}{%
  \Doi{10.1103/PhysRevD.71.124004}{\bibinfo {journal} {Phys.Rev.}}\ }%
  \textbf{\bibinfo {volume} {D71}},\ \bibinfo {pages} {124004} (\bibinfo {year}
  {2005}),\ \Eprint{http://arxiv.org/abs/gr-qc/0503044}{arXiv:gr-qc/0503044
  [gr-qc]}%
  \bibAnnoteFile{NoStop}{Blanchet:2005tk}%
\bibitem{Blanchet:2006zz}%
  \BibitemOpen
  \bibfield{author}{%
  \bibinfo {author} {\bibfnamefont{L.}~\bibnamefont{Blanchet}},\ }%
  \bibfield{journal}{%
  \bibinfo {journal} {Living Rev.Rel.}\ }%
  \textbf{\bibinfo {volume} {9}},\ \bibinfo {pages} {4} (\bibinfo {year}
  {2006})%
  \bibAnnoteFile{NoStop}{Blanchet:2006zz}%
\bibitem{Damour:2004bz}%
  \BibitemOpen
  \bibfield{author}{%
  \bibinfo {author} {\bibfnamefont{T.}~\bibnamefont{Damour}}, \bibinfo {author}
  {\bibfnamefont{A.}~\bibnamefont{Gopakumar}},\ and\ \bibinfo {author}
  {\bibfnamefont{B.~R.}\ \bibnamefont{Iyer}},\ }%
  \bibfield{journal}{%
  \Doi{10.1103/PhysRevD.70.064028}{\bibinfo {journal} {Phys.Rev.}}\ }%
  \textbf{\bibinfo {volume} {D70}},\ \bibinfo {pages} {064028} (\bibinfo {year}
  {2004}),\ \Eprint{http://arxiv.org/abs/gr-qc/0404128}{arXiv:gr-qc/0404128
  [gr-qc]}%
  \bibAnnoteFile{NoStop}{Damour:2004bz}%
\bibitem{Fujita:2010xj}%
  \BibitemOpen
  \bibfield{author}{%
  \bibinfo {author} {\bibfnamefont{R.}~\bibnamefont{Fujita}}\ and\ \bibinfo
  {author} {\bibfnamefont{B.~R.}\ \bibnamefont{Iyer}},\ }%
  \bibfield{journal}{%
  \Doi{10.1103/PhysRevD.82.044051}{\bibinfo {journal} {Phys. Rev.}}\ }%
  \textbf{\bibinfo {volume} {D82}},\ \bibinfo {pages} {044051} (\bibinfo {year}
  {2010}),\ \Eprint{http://arxiv.org/abs/1005.2266}{arXiv:1005.2266 [gr-qc]}%
  \bibAnnoteFile{NoStop}{Fujita:2010xj}%
\bibitem{Lackey:2011vz}%
  \BibitemOpen
  \bibfield{author}{%
  \bibinfo {author} {\bibfnamefont{B.~D.}\ \bibnamefont{Lackey}}, \bibinfo
  {author} {\bibfnamefont{K.}~\bibnamefont{Kyutoku}}, \bibinfo {author}
  {\bibfnamefont{M.}~\bibnamefont{Shibata}}, \bibinfo {author}
  {\bibfnamefont{P.~R.}\ \bibnamefont{Brady}},\ and\ \bibinfo {author}
  {\bibfnamefont{J.~L.}\ \bibnamefont{Friedman}}}%
   (\bibinfo {year} {2011}),\ \bibinfo {note} {* Temporary entry *},\
  \Eprint{http://arxiv.org/abs/1109.3402}{arXiv:1109.3402 [astro-ph.HE]}%
  \bibAnnoteFile{NoStop}{Lackey:2011vz}%
\bibitem{Bernuzzi:2011aj}%
  \BibitemOpen
  \bibfield{author}{%
  \bibinfo {author} {\bibfnamefont{S.}~\bibnamefont{Bernuzzi}}, \bibinfo
  {author} {\bibfnamefont{A.}~\bibnamefont{Nagar}},\ and\ \bibinfo {author}
  {\bibfnamefont{A.}~\bibnamefont{Zenginoglu}},\ }%
  \bibfield{journal}{%
  \Doi{10.1103/PhysRevD.84.084026}{\bibinfo {journal} {Phys.Rev.}}\ }%
  \textbf{\bibinfo {volume} {D84}},\ \bibinfo {pages} {084026} (\bibinfo {year}
  {2011}),\ \Eprint{http://arxiv.org/abs/1107.5402}{arXiv:1107.5402 [gr-qc]}%
  \bibAnnoteFile{NoStop}{Bernuzzi:2011aj}%
\bibitem{Damour:1997ub}%
  \BibitemOpen
  \bibfield{author}{%
  \bibinfo {author} {\bibfnamefont{T.}~\bibnamefont{Damour}}, \bibinfo {author}
  {\bibfnamefont{B.~R.}\ \bibnamefont{Iyer}},\ and\ \bibinfo {author}
  {\bibfnamefont{B.}~\bibnamefont{Sathyaprakash}},\ }%
  \bibfield{journal}{%
  \Doi{10.1103/PhysRevD.57.885}{\bibinfo {journal} {Phys.Rev.}}\ }%
  \textbf{\bibinfo {volume} {D57}},\ \bibinfo {pages} {885} (\bibinfo {year}
  {1998}),\ \Eprint{http://arxiv.org/abs/gr-qc/9708034}{arXiv:gr-qc/9708034
  [gr-qc]}%
  \bibAnnoteFile{NoStop}{Damour:1997ub}%
\bibitem{Finn:2000sy}%
  \BibitemOpen
  \bibfield{author}{%
  \bibinfo {author} {\bibfnamefont{L.~S.}\ \bibnamefont{Finn}}\ and\ \bibinfo
  {author} {\bibfnamefont{K.~S.}\ \bibnamefont{Thorne}},\ }%
  \bibfield{journal}{%
  \bibinfo {journal} {Phys. Rev. D}\ }%
  \textbf{\bibinfo {volume} {62}},\ \bibinfo {pages} {124021} (\bibinfo {year}
  {2000}),\ \Eprint{http://arxiv.org/abs/gr-qc/0007074}{gr-qc/0007074}%
  \bibAnnoteFile{NoStop}{Finn:2000sy}%
\bibitem{Barack:2007tm}%
  \BibitemOpen
  \bibfield{author}{%
  \bibinfo {author} {\bibfnamefont{L.}~\bibnamefont{Barack}}\ and\ \bibinfo
  {author} {\bibfnamefont{N.}~\bibnamefont{Sago}},\ }%
  \bibfield{journal}{%
  \Doi{10.1103/PhysRevD.75.064021}{\bibinfo {journal} {Phys.Rev.}}\ }%
  \textbf{\bibinfo {volume} {D75}},\ \bibinfo {pages} {064021} (\bibinfo {year}
  {2007}),\ \Eprint{http://arxiv.org/abs/gr-qc/0701069}{arXiv:gr-qc/0701069
  [gr-qc]}%
  \bibAnnoteFile{NoStop}{Barack:2007tm}%
\bibitem{Akcay:2010dx}%
  \BibitemOpen
  \bibfield{author}{%
  \bibinfo {author} {\bibfnamefont{S.}~\bibnamefont{Akcay}},\ }%
  \bibfield{journal}{%
  \Doi{10.1103/PhysRevD.83.124026}{\bibinfo {journal} {Phys.Rev.}}\ }%
  \textbf{\bibinfo {volume} {D83}},\ \bibinfo {pages} {124026} (\bibinfo {year}
  {2011}),\ \Eprint{http://arxiv.org/abs/1012.5860}{arXiv:1012.5860 [gr-qc]}%
  \bibAnnoteFile{NoStop}{Akcay:2010dx}%
\bibitem{Teukolsky:1973}%
  \BibitemOpen
  \bibfield{author}{%
  \bibinfo {author} {\bibfnamefont{S.~A.}\ \bibnamefont{Teukolsky}},\ }%
  \bibfield{journal}{%
  \bibinfo {journal} {Astrophys. J.}\ }%
  \textbf{\bibinfo {volume} {185}},\ \bibinfo {pages} {635} (\bibinfo {year}
  {1973})%
  \bibAnnoteFile{NoStop}{Teukolsky:1973}%
\bibitem{Teukolsky:1974}%
  \BibitemOpen
  \bibfield{author}{%
  \bibinfo {author} {\bibfnamefont{S.~A.}\ \bibnamefont{Teukolsky}}\ and\
  \bibinfo {author} {\bibfnamefont{W.~H.}\ \bibnamefont{Press}},\ }%
  \bibfield{journal}{%
  \bibinfo {journal} {Astrophys. J.}\ }%
  \textbf{\bibinfo {volume} {193}},\ \bibinfo {pages} {443} (\bibinfo {year}
  {1974})%
  \bibAnnoteFile{NoStop}{Teukolsky:1974}%
\bibitem{Bernuzzi:2010ty}%
  \BibitemOpen
  \bibfield{author}{%
  \bibinfo {author} {\bibfnamefont{S.}~\bibnamefont{Bernuzzi}}\ and\ \bibinfo
  {author} {\bibfnamefont{A.}~\bibnamefont{Nagar}},\ }%
  \bibfield{journal}{%
  \Doi{10.1103/PhysRevD.81.084056}{\bibinfo {journal} {Phys. Rev.}}\ }%
  \textbf{\bibinfo {volume} {D81}},\ \bibinfo {pages} {084056} (\bibinfo {year}
  {2010}),\ \Eprint{http://arxiv.org/abs/1003.0597}{arXiv:1003.0597 [gr-qc]}%
  \bibAnnoteFile{NoStop}{Bernuzzi:2010ty}%
\bibitem{Bernuzzi:2010xj}%
  \BibitemOpen
  \bibfield{author}{%
  \bibinfo {author} {\bibfnamefont{S.}~\bibnamefont{Bernuzzi}}, \bibinfo
  {author} {\bibfnamefont{A.}~\bibnamefont{Nagar}},\ and\ \bibinfo {author}
  {\bibfnamefont{A.}~\bibnamefont{Zenginoglu}},\ }%
  \bibfield{journal}{%
  \Doi{10.1103/PhysRevD.83.064010}{\bibinfo {journal} {Phys.Rev.}}\ }%
  \textbf{\bibinfo {volume} {D83}},\ \bibinfo {pages} {064010} (\bibinfo {year}
  {2011}),\ \Eprint{http://arxiv.org/abs/1012.2456}{arXiv:1012.2456 [gr-qc]}%
  \bibAnnoteFile{NoStop}{Bernuzzi:2010xj}%
\bibitem{Bernuzzi:2012}%
  \BibitemOpen
  \bibfield{author}{%
  \bibinfo {author} {\bibfnamefont{S.}~\bibnamefont{Bernuzzi}}, \bibinfo
  {author} {\bibfnamefont{A.}~\bibnamefont{Nagar}},\ and\ \bibinfo {author}
  {\bibfnamefont{A.}~\bibnamefont{Zenginoglu}}}%
   (\bibinfo {year} {2012})%
  \bibAnnoteFile{NoStop}{Bernuzzi:2012}%
\end{thebibliography}%

\end{document}